\documentclass{jov}

\usepackage[utf8]{inputenc}
\usepackage[T1]{fontenc}
\usepackage{lmodern}
\usepackage{amsmath}
\usepackage{amsfonts}
\usepackage{amssymb}
\usepackage{tabularx}
\usepackage{natbib}
\usepackage{gensymb}
\usepackage{units}
\usepackage{eurosym}
\usepackage[english]{babel}
\usepackage[normalem]{ulem}
\usepackage{float}

\usepackage{lineno}

\PassOptionsToPackage{pdftex}{graphicx}
\usepackage{graphicx} 
\usepackage[dvipsnames]{xcolor}

\usepackage{xcolor}
\definecolor{orange}{rgb}{.5,.35,0}
\definecolor{darkgreen}{rgb}{0,0,0}

\newcommand{\daniel}[1]{{\color{black}#1}}
\newcommand{\rev}[1]{{\color{darkgreen}#1}}
\newcommand\del{\bgroup\markoverwith{\textcolor{red}{\rule[0.5ex]{2pt}{4.0pt}}}\ULon}

\usepackage{float}

\begin{document}

\title{\daniel{Task-dependence in scene perception: Head unrestrained viewing using mobile eye-tracking.}}

\abstract{
Real-world scene perception is typically studied in the laboratory using static picture viewing with restrained head position. Consequently, the transfer of results obtained in this paradigm to real-word scenarios has been questioned. The advancement of mobile eye-trackers and the progress in image processing, however, permit a more natural experimental setup that, at the same time, maintains the high experimental control from the standard laboratory setting. We investigated eye movements while participants were standing in front of a projector screen and explored images under four specific task instructions. Eye movements were recorded with a mobile eye-tracking device and raw gaze data was transformed from head-centered into image-centered coordinates. We observed differences between tasks in temporal and spatial eye-movement parameters and found that the bias to fixate images near the center differed between tasks. Our results demonstrate that current mobile eye-tracking technology and a highly controlled design support the study of fine-scaled task dependencies in an experimental setting that permits more natural viewing behavior than the static picture viewing paradigm.
}

\author{Backhaus}{Daniel}
 {Experimental and Biological Psychology}
 {University of Potsdam, Potsdam, Germany}
 {http://}{lars.rothkegel@uni-potsdam.de}

\author{Engbert}{Ralf}
 {Experimental and Biological Psychology}
 {University of Potsdam, Potsdam, Germany}
 {http://}{ralf.engbert@uni-potsdam.de}
 
 \author{Rothkegel}{Lars O. M.}
 {Experimental and Biological Psychology}
 {University of Potsdam, Potsdam, Germany}
 {http://}{lars.rothkegel@uni-potsdam.de}
 
\author{Trukenbrod}{Hans A.}
 {Experimental and Biological Psychology}
 {University of Potsdam, Potsdam, Germany}
 {http://}{hans.trukenbrod@uni-potsdam.de}


\maketitle

\newpage

\section{Introduction}
Over the course of the last decades, scene viewing has been used to study the allocation of attention on natural images. In recent years, however, several limitations of the paradigm have been criticized and a paradigmatic shift towards real-world scenarios has been suggested \citep[e.g.,][]{tatler2011}. Here, we propose a different approach that gradually moves from scene viewing towards more natural tasks. This provides a link between the two opposing approaches and helps to understand to which degree eye-movement behavior generalizes across tasks.

In the scene-viewing paradigm, eye movements are recorded in the laboratory from participants looking at an image for a few seconds on a computer screen \citep{henderson2003,Rayner.QJExpPsychol.2009}. Usually, participants get an unspecific instruction to view the image (``free viewing'') or alternatively to memorize the image for a subsequent recall test. In most experiments, images consist of color photographs of the real world selected by the experimenter. As a consequence, within and between experiments images differ considerably with respect to their low-level features (color, edges), features at more complex levels (shapes, objects, 3D arrangement) and their high-level features \citep[semantic category, action affordances;][]{Malcolm.TrendsCognSci.2016}.  

One reason why scene viewing has become an intensively used paradigm is that it allows \daniel{researchers } to study eye movements and, hence, the overt allocation of attention on ecologically valid, complex stimuli under highly-controlled laboratory conditions. Since the mapping of the eye position to coordinates within an image is straightforward, much research has focused on the question \daniel{of} image-features influence on eye movements in a bottom-up fashion, that is, independent of the internal state of the observer. Examples of correlations between simple low-level features and fixation positions are local luminance contrast and edge density \citep{Mannan.Perception.1997,Reinagel.NetworkCompNeural.1999,Tatler.VisionRes.2005}. But the correlations are not limited to low-level image features. More complex high-level features that correspond to shapes and objects improve predictions substantially \cite[e.g., faces, persons, cars;][]{Cerf.AdvNeuralInfoProcSyst.2007,Einhauser.JVis.2008,Judd.CompVision.2009}. The idea of bottom-up selection of fixation locations based on image features led to the development of saliency models \citep{koch1985,itti2001} and a large variety of models has been put forward \citep[e.g.,][]{Bruce.JVis.2009,kummerer2016,Parkhurst:2002jy}. In particular with the development of sophisticated machine-learning algorithms, these models predict fixation locations well when evaluated with a data set obtained under the free viewing instruction \citep{mit-saliency-benchmark}. Beside their influence on fixation locations, both low-level and high-level image features have also been shown to influence fixation durations \citep{nuthmann2017,tatler2017}.

Already in their anecdotal works, \cite{buswell1935} and \cite{yarbus1967eye} demonstrated that eye-movement patterns depend on the instruction given to the viewer and not just the bottom-up appearance of an image. This top-down influence has often been replicated since \citep{castelhano2009,DeAngelus:2009bm,Mills:2011}. Furthermore, in paradigms where participants pursue a specific natural task like preparing a sandwich \citep{Hayhoe:2003} or making a cup of tea \citep{land1999}, the necessities of motor actions dominate eye-movement behavior. Here, eye movements support task execution by bringing critical information to the foveal region \emph{just-in-time} \citep{Ballard.BehavBrainSci.1997,Land.BOOK.2009} or as look-ahead fixations on objects needed later during a task \citep{Pelz.VisionRes.2001}. Similar conclusions have been made for various other activities like driving \citep{Land:2001dl}, cycling \citep{Vansteenkiste.PLoSOne.2014}, walking \citep{Matthis.CurrBiol.2018,Rothkopf.JVis.2007}, and ball games \citep{Land.NatNeurosci.2000,Land.PhilosTRoySocB.1997}. To align the bottom-up approach with the contradictory findings of top-down control, it is often implicitly assumed that scene viewing without specific instruction provides the means to isolate task-free visual processing. It is a default mode of viewing that can be overridden by the presence of specific tasks. But it is more likely that participants chose a task based on their internal agenda and researchers are simply unaware of the chosen task in the free viewing condition \citep{tatler2011}.

In addition, \cite{tatler2011} criticized several limitations of the scene-viewing paradigm. Participants are seated in front of a computer screen with their head on a chin-rest and are asked to minimize head and body movements. Images are presented for a few seconds after a sudden onset on a computer screen limiting the field of view to the size of the display. The viewpoint is fixed by the photographer and contains compositional biases \citep{Tatler.VisionRes.2005}. A situation that substantially differs from our experience in daily life, where we are free to move, where scenes emerge slowly (e.g., by opening a door) and our binocular field of view encompasses 200\degree--220\degree~of visual angle \citep{Loschky.JVis.2017,Roenne.1915}. As a consequence, visual processing and reconstruction of image content might \daniel{differ a lot} during scene viewing \daniel{and} in real-world tasks as some depth cues (stereo and motion parallax) and motion cues (both egomotion and external motion) are missing in static images. Furthermore, scene viewing utilizes only a portion of the repertoire of eye-movement behaviors needed for other tasks. For example, participants typically make smaller gaze shifts during scene viewing than in everyday activities \citep{Land:2001hl}. This is at least in part generated by the restrictions of the task, since saccade amplitudes scale with image size \citep{vonWartburg:2007} and large gaze shifts are usually supported by head movements \citep{Gossens.ExpBrainRes.1997, Stahl:1999kp}, but \daniel{in the classical scene-viewing setup these head movements are suppressed.} Hence, \cite{tatler2011} suggested to put a stronger emphasis on the study of eye guidance in natural behavior.

Only few studies have directly compared viewing behavior under similar conditions in the real world and in the laboratory. As an exception, \cite{tHart:2009kj} recorded eye movements during free exploration of various in- and outdoor environments using a mobile eye-tracker. In a second session the recorded head-centered videos were replayed in the laboratory either as a continuous video or randomly chosen frames from the video were presented for 1~s as in the scene-viewing paradigm. Interobserver consistency was highest when observers viewed static images. The result could partially be explained by a bias to fixate near the center, which was strongest in the static image condition as initial fixations are typically directed towards the image center after a sudden onset \citep[cf.,][]{rothkegel2017,tatler2007}. In addition, during free exploration fixation locations showed a greater vertical variability as participants also looked down on the path while moving forward \citep[cf.,][]{t_hart2012}. Finally, fixations during free exploration were better predicted by fixations from the replay condition than the static image condition, demonstrating that the scene-viewing paradigm has only limited explanatory power for eye movements during free exploration. In a follow-up experiment, \cite{Foulsham.CJExpPsychol.2017} demonstrated that keeping the correct order of images in the static image condition changes gaze patterns and improves the predictability of fixation locations during free exploration. But this prediction was no better than just a general bias to fixate near the center independent of image content. In a similar vein, \cite{Foulsham:2011hs} compared eye movements while navigating on a campus with eye movements while watching the head-centered videos. Both conditions showed a strong bias to fixate centrally. However, during walking gaze was shifted slightly below the horizon, while gaze was shifted slightly above the horizon during watching. Furthermore, while walking participants spent more time looking at the near path, less time on distant objects and pedestrians were less likely fixated when they approached the observer in line with the observation that social context modulates the amount of gaze directed towards real people \citep{Laidlaw.PNAS.2011,Risko.CurrDirPsycholSci.2016}.

It is not surprising that eye guidance during scene viewing strongly differs from other natural tasks given the limited overlap of tasks and environments. Even in studies that sought to directly compare laboratory and real-world behavior \citep{Foulsham:2011hs,Dicks:2010be,tHart:2009kj}, several aspects differed between conditions (e.g., size of field of view, task affordances). While scene viewing cannot be thought of as a proxy for eye movements in natural tasks, a paradigmatic shift away from scene viewing might be premature. For several reasons we advocate for a line of research that makes a smooth transition from the classical scene-viewing paradigm towards more natural tasks. First, the scene-viewing paradigm deals with important aspects of our daily lifes as people are constantly engaged in viewing static scenes. Second, the extensive research on scene viewing provides a solid theoretical basis for future research and has led to the development of computational models that predict scanpaths  \citep{engbert2015,lemeur2015,schutt2017} and fixation durations \citep{nuthmann2010,tatler2017}. Third, due to the advancement of mobile eye-trackers, it is technically straightforward to address limitations of the paradigm \citep{tatler2011}, while keeping the benefits of the highly controlled experimental conditions in the laboratory. Fourth, eye guidance in scene viewing is not decoupled from other tasks as some behaviors generalize to other domains. For example, the observation of the central fixation bias \citep{tatler2007}, that is, the tendency of viewers to place fixations near the center of an image, has been observed in natural tasks like walking, tea making, and card sorting \citep{tHart:2009kj,Foulsham:2011hs,Ioannidou.JEMR.2016}. Finally, the scene-viewing paradigm provides a fruitful test bed for theoretical assumptions about eye guidance derived from other paradigms \citep[e.g., inhibition of return;][]{rothkegel2016,smith2009} and can advance the development of theories of eye guidance in general.

\daniel{We} suggest to adjust the scene-viewing paradigm \daniel{step-by-step} to deal with its limitations. This approach allows \daniel{researchers} to systematically investigate the influence of individual factors. In this study, we remove \daniel{some} limitations of the paradigm while keeping high overall eye-tracking accuracy. In contrast to the classical scene viewing paradigm\daniel{, in our experiment} participants stood in front of a projector screen and viewed images with a specific instruction. \daniel{Other experimental aspects (e.g., size of field of view, color stimulus material, sudden image onset, possible interactions with the stimulus material) were kept to stay comparable to the classical scene-viewing setup.} Eye movements were recorded with a mobile eye-tracker and participants were free to make body and head movements. \daniel{Note that we did not \rev{encourage} large-scale head or body movements or force participants to move in front of the screen. But without being explicit, we reduced participants' restrictions and gave viewers the possibility to move.}

\rev{The main purpose of our study was to investigate whether established task differences can be reproduced reliably under relaxed viewing condition. For example, a possible body-posture related modulation of image-independent fixation tendencies could override task differences that were observed in earlier studies. Thus, the key contribution of study is to demonstrate the stability of task effects under more natural viewing conditions.

If task effects turn out to be reliable in our paradigm, we expect to find differences in basic eye-movement parameters as in the classical scene-viewing paradigm, e.g.,} shorter fixation durations and longer saccade amplitudes for search tasks \citep{Mills:2011, castelhano2009}. For fixation locations, we expected a more extended range of fixation locations for search tasks \citep{tatler2007}. For the central fixation bias the artificial situation in the laboratory (e.g., sudden image onset, \citep{rothkegel2017, tatler2011} can partly explain the tendency to fixate images near the image center. We expected modulation of the central fixation bias by task since search behavior will typically lead to a broader distribution of fixation locations.

\rev{In the following section, we describe our methods, where we outline the processing pipeline to check data quality under this setup and how to convert gaze recorded by a mobile eye-tracker into image coordinates. Next, we report our main results, report an early task-independent central fixation bias and a late task-dependent central fixation bias. We continue with analyses of basic eye-movement parameters like fixation durations, saccade amplitudes, and distribution of fixation locations across tasks. Finally, we  investigate how well fixation locations from one task predict fixation locations from another task in our relaxed setup. We close with a discussion.}

\section{Methods}
\subsection{Participants}

For this study, we \rev{used data of} 32 students of the University of Potsdam with normal or corrected to normal vision. On average participants were 22.8 years old (18--36 years) and 31 participants were female. Participants received credit points or a monetary compensation of 10\euro. To increase compliance with the task, we offered participants an additional incentive of up to 3\euro~for correctly answering questions after each image (in sum 60~questions). The work was carried out in accordance with the Declaration of Helsinki. Informed consent was obtained for experimentation by all participants.

\subsection{Stimulus Presentation, Laboratory Setup \& Procedure}

Participants were instructed to look at images while standing in front of a 110-inch projector screen at a viewing distance of 270~cm. Images were projected with a luminance calibrated video beamer (JVC -- DLA-X9500B; frame rate 60~Hz, resolution 1920$\times$1080 pixels; Victor Company of Japan, Limited, JVC, Yokohama, Japan). Eye movements were recorded binocularly using the SMI Eye-Tracking Glasses (SMI-ETG 2W, SensoMotoric Instruments, Teltow, Germany) with a sampling rate of 120~Hz. In addition, the scene camera of the Eye-Tracking Glasses recorded the field of view of the participant with a resolution of 960$\times$720 pixels (60\degree$\times$46\degree~of visual angle) at 30~Hz.

All images were presented with a resolution of 1668$\times$828 pixels at the center of the screen. Images were embedded in a gray frame with QR-markers ($126 \times 126$ pixels; cf., Fig.~\ref{fig:Transformation}) and covered $40.6^\circ$~of visual angle in the horizontal and $20.1^\circ$~in the vertical dimension. Images were colored scene photographs taken by the authors, \rev{every single image} contained zero to ten humans and zero to ten animals. \rev{We used 27 images with people and animals, one image with only animals, one image with only people and one image with neither people nor animals.} Furthermore, images were selected by having an overall sharpness, were taken in different countries and did not contain prominent text. \rev{We selected 30 images. Each image could appear in every condition and was presented in two conditions to every single participant.}

The experiment consisted of four blocks. In each block, participants viewed images under one of four instructions. Under two instructions, participants had to count the number of people (Count People) or count the number of animals in an image (Count Animals). Under the two remaining instructions, participants had to guess the time of day when an image was taken (Guess Time) and guess the country in which an image was taken (Guess Country). We expected the count instructions to resemble search tasks, since the entire image had to be thoroughly examined to give a correct answer, while the guess instructions were thought to resemble the free viewing instruction but with a stronger focus on one aspect of the image for all participants. In each block, we presented 15 images for 8~seconds. While the order of instructions was counterbalanced across participants, each image was randomly assigned to two of the four instructions. 

\begin{figure}[H]
\begin{center}
\includegraphics[width =.9\textwidth]{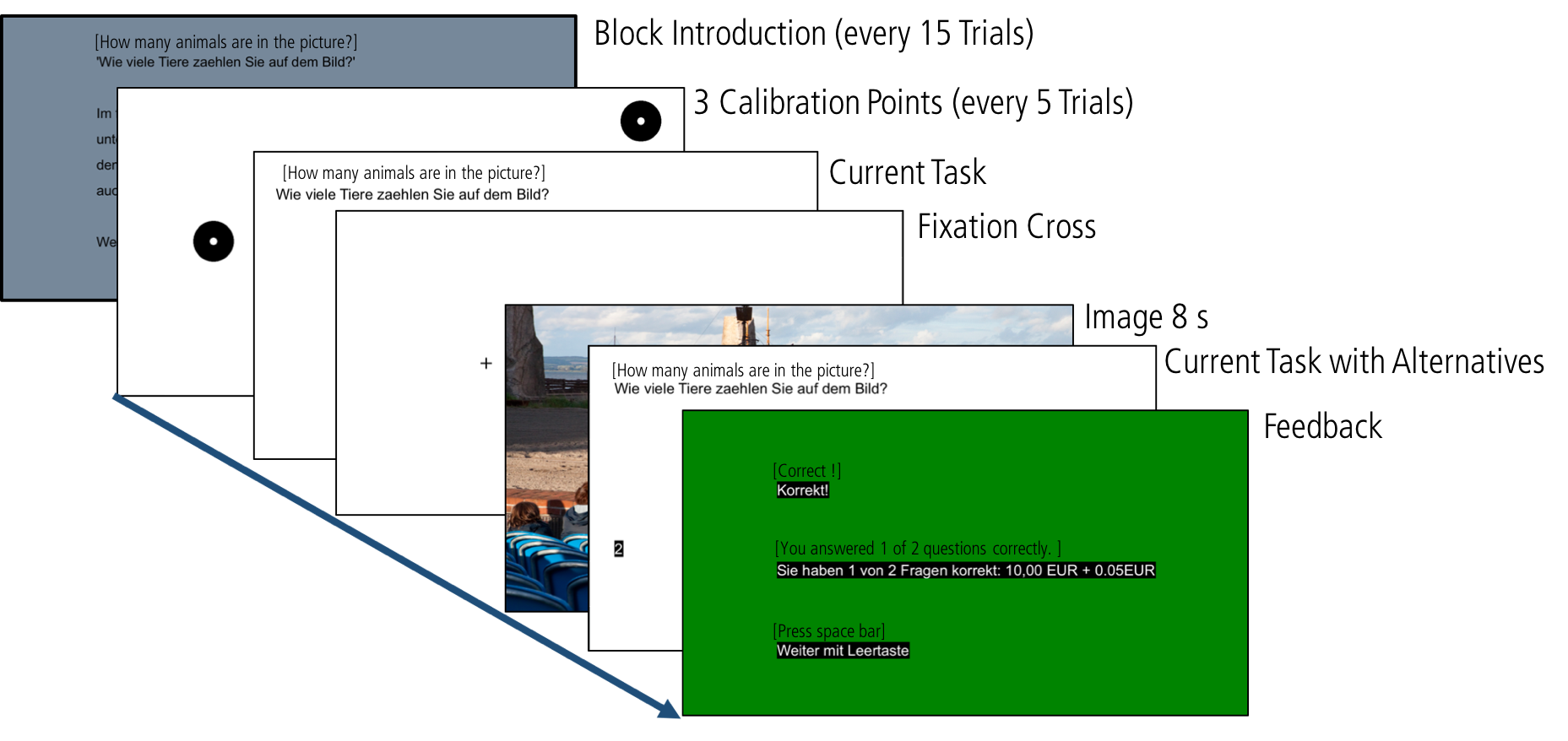}
\caption{Sequence of events in the scene-viewing experiment. \label{fig:Procedure}}
\end{center}
\end{figure}

At the beginning of each block we presented a detailed instruction for the upcoming task, followed by a three point calibration (Fig.~\ref{fig:Procedure}). Individual trials began with a 1-second reminder of the instruction, followed by a black fixation cross ($0.73\degree \times 0.73\degree$) presented on a white background for 3 seconds. Participants were instructed to fixate the fixation cross until the image appeared. Fixation crosses appeared on a grid of 15 fixed positions: three vertical positions (25~\%, 50~\% and 75~\% of the projector screens vertical size) and five horizontal positions (20~\%, 35~\%, 50~\%, 65~\% and 80~\% of the projector screens horizontal size). Afterwards participants were free to explore the image for 8 seconds. At the end of a trial, participants had to answer orally a multiple choice question with three alternatives presented on the screen. We gave immediate feedback and each correct answer was rewarded with 0.05\euro. The instructor pressed a button to continue with the next trial, which started with a brief reminder of the instruction. The eyes were calibrated at the beginning of each block and after every fifth image. In addition, instructors could force a new calibration after a trial if fixations deviated more than $\sim$1\degree from the fixation cross during the initial fixation check.  

\subsection{Raw Data Processing}

\subsubsection{Transformation}
\begin{figure}[H]
\begin{center}
\includegraphics[width = .9\textwidth]{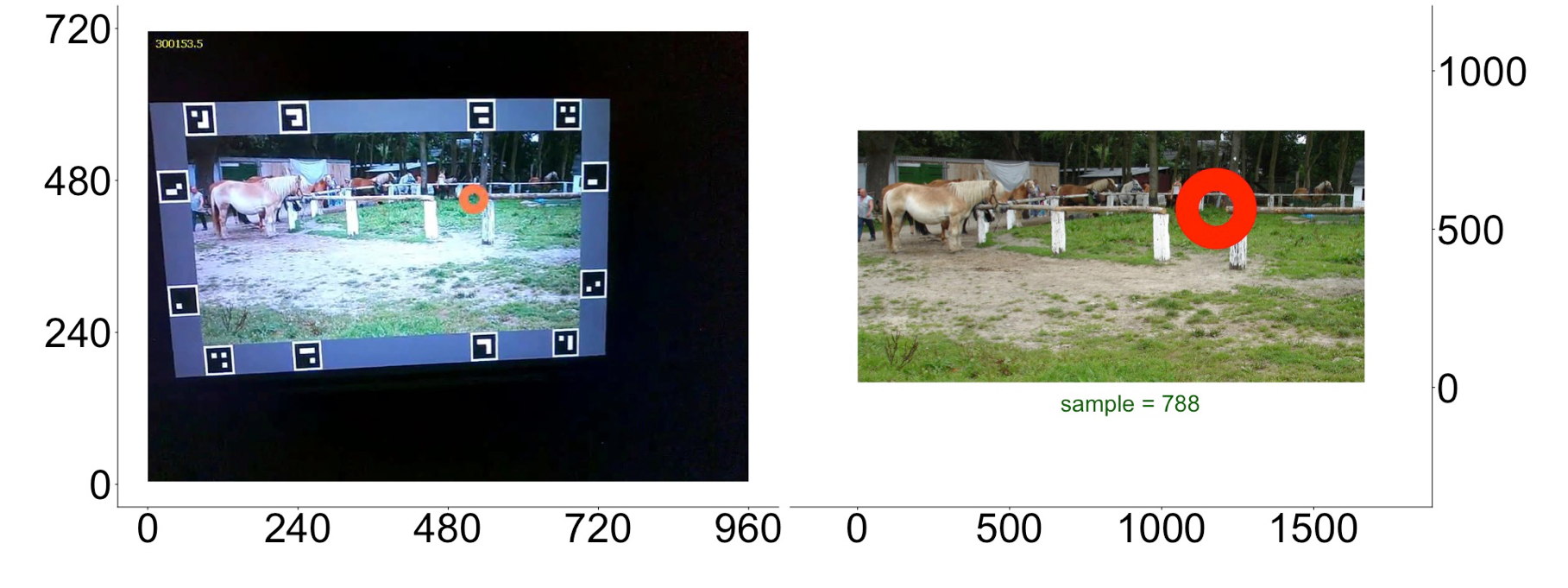}
\caption{\rev{Transformation of scene-camera coordinates (subpixel level) into image coordinates in pixels. Left panel: Frame taken by SMI ETG-120Hz scene camera with measured fixation location (circle). Right panel: The same frame and fixation in image coordinates.}}
\label{fig:Transformation}
\end{center}
\end{figure}

The experimentally measured eye-positions were given in coordinates of the scene camera of the mobile eye-tracker. Thus, raw data \rev{subpixel (1/100 pixel)} values had to be transformed into coordinates of the presented image (Fig.~\ref{fig:Transformation}). To achieve this, we used a projective transformation provided by the computer vision toolbox in the MATLAB programming language (MATLAB 2015b, The MathWorks, Natick/MA, USA). The required locations of image corners were extracted from the scene-camera output frame by frame, using 12 unique QR-markers, which were presented around the images. Automatic QR-marker detection as well as detection of image corners were done with the Offline Surface Tracker module of the Pupil Labs software Pupil Player \citep{KassnerPateraBulling:2014}. To synchronize the time of both devices, we sent UDP-messages from the presentation-computer to the recording unit of the eye tracker. \rev{As a result of this calculation we worked with three trajectories in image coordinates: Two monocular data streams and one binocular data stream. First, saccade detection was performed with both monocular eye-data streams (see next section). Second, we calculated mean fixation positions based on the binocular eye-data stream (note that the binocular data are not the simple mean of both monocular trajectories). Pilot analyses of the fixation positions indicated higher reliability of the binocular position estimate compared to averaging of monocular positions. }

\subsubsection{Saccade detection}
For saccade detection we applied a velocity-based algorithm \citep{engbert2003,engbert2006}. The algorithm marks all parts of an eye trajectory as a saccade that have a minimum amplitude of 0.5$^\circ$~and exceed a velocity threshold for at least 3 successive data samples (16.7 ms). The velocity threshold is computed as a multiple $\lambda$ of the median-based standard deviation of the eye-trajectories' velocity during a trial. We carried out a systematic analysis with varying threshold multipliers $\lambda$ to identify detection parameters for obtaining robust results \citep{engbert:2016}. Here, we computed the velocity threshold with a multiplier $\lambda = 8$. We first analyzed both monocular eye-trajectories to identify potential saccades and kept all binocular events. 

Following \cite{Hessels:2018jj}, it is important to clearly define what a fixation means in the context of a specific analysis. In the current work, fixations refer to moments of relative stability on an image, regardless of eye-in-head and body movements. Fixations were computed as the epoch between two subsequent saccades. The binocular eye-data stream provided from the recording unit was \daniel{transformed} and used to calculate the mean fixation position.

\subsection{Data quality}
\subsubsection{Raw data quality}
\rev{In total, we recruited 42 participants to get our planed 32 participants.} Five participants had to be replaced as the experimenter was not able to calibrate them reliably \rev{(these participants did not finish the experiment)}. Another five participants had to be replaced since at least a fifth of their data was missing due to blinks and low data quality (see next paragraph). 
 
To ensure high data quality, we marked blinks and epochs with high noise in the eye trajectories. For the detection of blinks, we made use of the blink detection provided by the SMI-ETG 2W. All fixations and saccades that contained a blink as well as all fixations and saccades with a blink during the preceding or succeeding event were removed from further analyses. Several other criteria were applied to detect unreliable events. First, we detected instable fixations (e.g. due to a strong jitter in the signal of the eye trajectory) by calculating the mean 2D standard deviation of the eye trajectory of all fixations. All fixations that contained epochs that exceeded the 2D standard deviation by a factor of 15 were removed from further analyses. Second, as saccades are stereotyped and ballistic movements, all saccades with a duration of more than 250~ms (30 samples) were removed. These saccades would be expected to have amplitudes, which go far beyond the dimensions of the projector screen; further we removed all saccades with amplitudes greater or equal to 25\degree. Third, we removed fixations located outside the image coordinates and fixations with a duration of less than 25~ms as well as with durations of more than 1000~ms. As a final criterion, we calculated the absolute deviation of participants' eye positions from the initial fixation cross. We computed the median deviation of the last 200~ms before the appearance of an image. Since we were not able to cancel the next trial and to immediately recalibrate with our setup, we removed trials with an absolute deviation greater than 2\degree. Overall, 40,182 fixations ($\sim$81\%  of 49,371) and 37,726 saccades ($\sim$80\%  of 47,425) remained for further analyses.

\subsubsection{Main sequence of saccade amplitude and peak velocity}
\begin{figure}[H]
\begin{center}
\includegraphics[width = .3\textwidth]{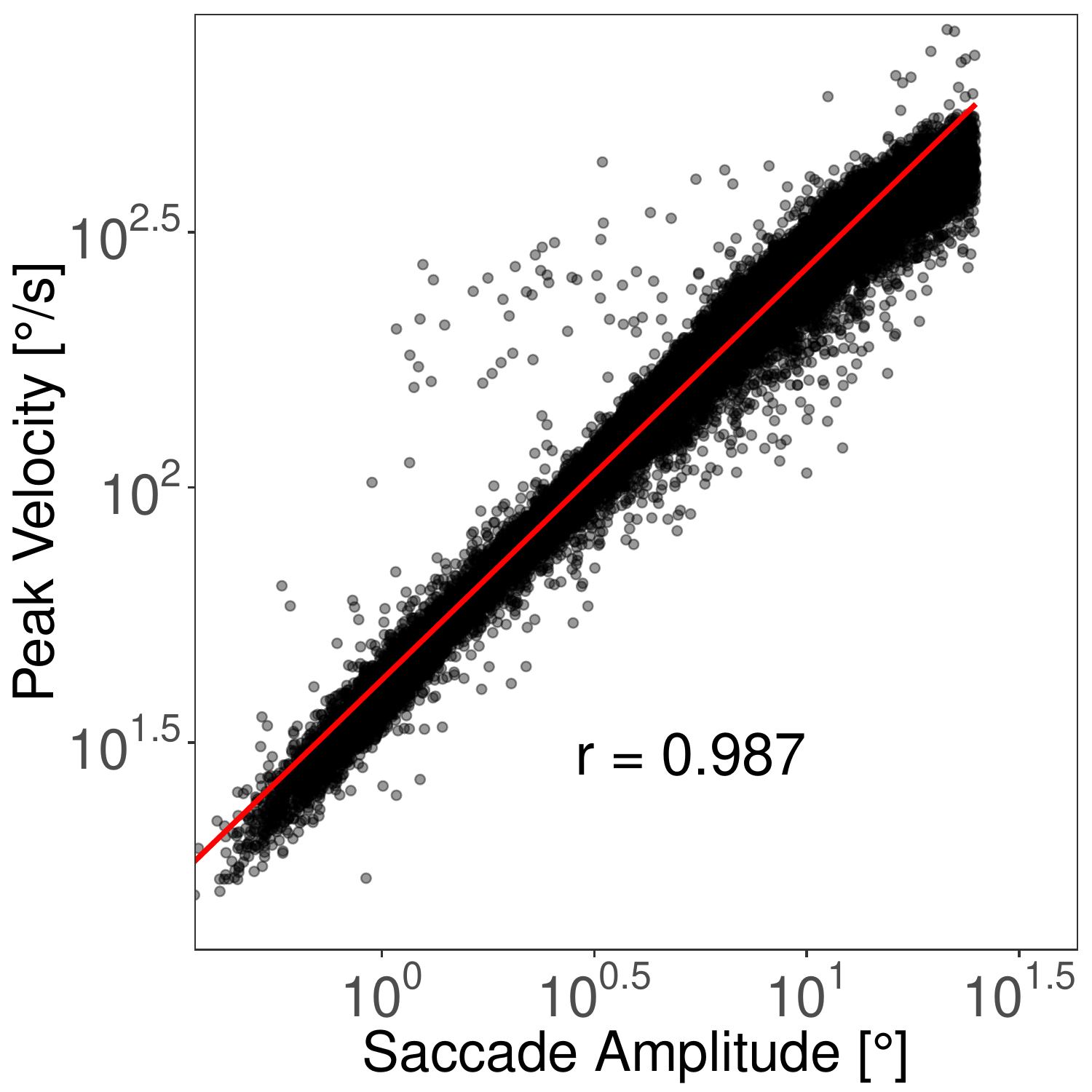}
\caption{Main sequence. Double-logarithmic representation of saccade amplitude and saccade peak velocity.}
\label{fig:MainSequence}
\end{center}
\end{figure}

Since saccades are stereotyped and ballistic movements, there is a high correlation between a saccade's amplitude and its peak velocity. We investigated this relationship by computing the main sequence, i.e., the double-logarithmic linear relation between saccade amplitude and peak velocity \citep{bahill1975main}. The 37,726 saccades in our data set range from about 0.5\degree~to about 25\degree~of visual angle, \daniel{due to our exclusion criteria} (Fig.~\ref{fig:MainSequence}). There is a strong linear relation in the main sequence with a very high correlation r=.987. Hence, the detected saccades behave as expected and were used for further analyses.

\begin{figure}[H]
\begin{center}
\color{black}
\includegraphics[width = .6\textwidth]{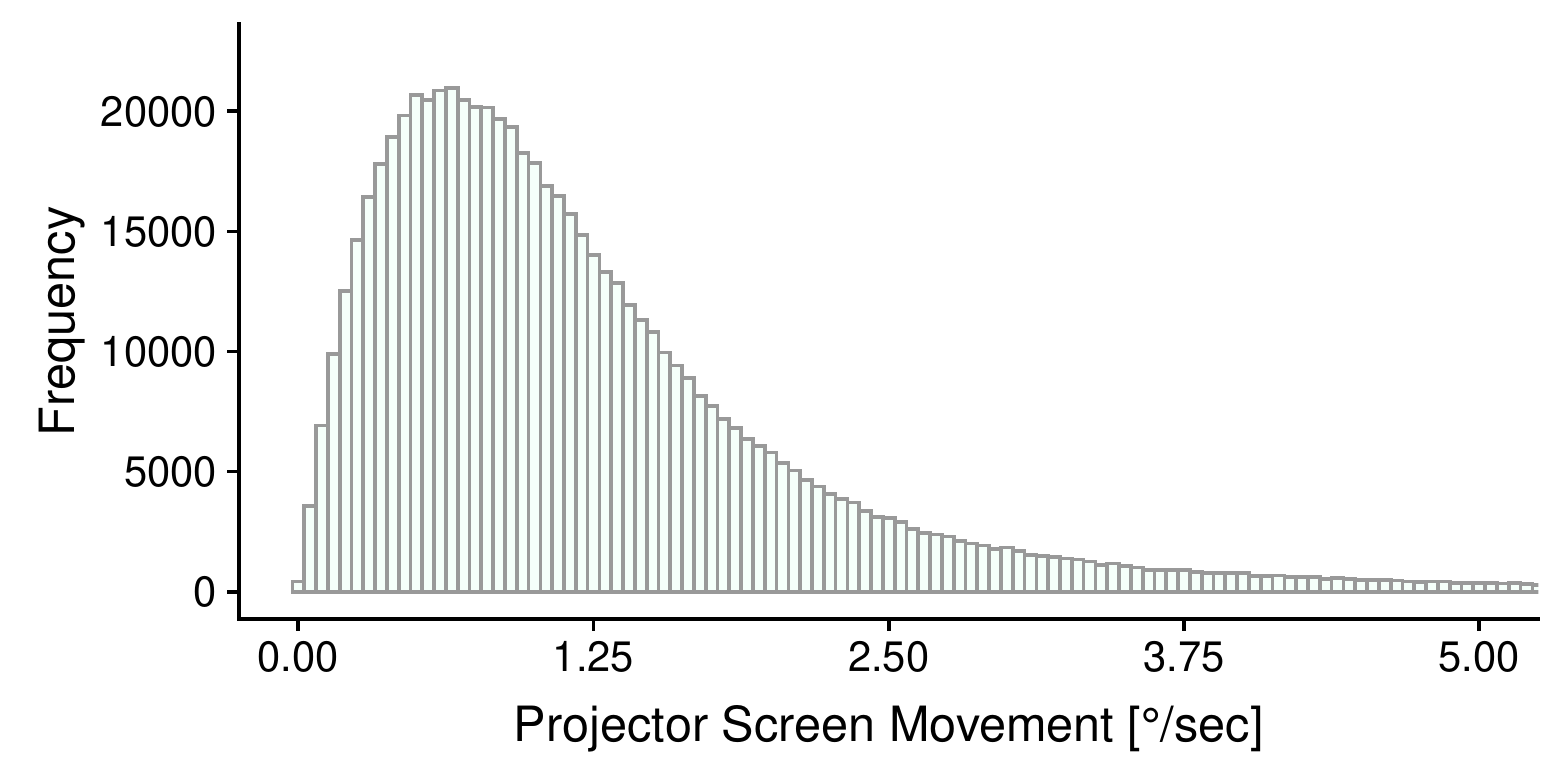}
\caption{Projector screen movement. As an approximation of head movements the projector screen movement is measured by tracking the position of QR-markers in the scene-camera video.}
\label{fig:ProjectorMovement}
\end{center}
\end{figure}

\daniel{\subsubsection{Head and body movements}

We realized a more natural body posture by recording without a chin rest and thereby enabling for small body and head movements in front of a projector screen. Even so, we did not expect large-scale head or body movements, as we did not \rev{encourage} gestures or movements explicitly in our tasks \citep{Epelboim:1995vc}. For an approximating measure of participants' movements in front of the screen, we made use of the QR-markers presented around the images. By tracking the marker positions in the scene-camera video, we receive a measure of participants' head position and angle relative to the projector screen. Figure~\ref{fig:ProjectorMovement} shows the distribution of the projector screen movements as an approximation for head and body movements. The distribution has a peak at around 1~$\degree/sec$ and only  few samples with velocities $\geq 2.5~\degree/sec.$ Thus, the majority of values do not exceed the velocities of fixational eye-movements. 
}

\subsubsection{Accuracy of the eye position}
\begin{figure}[H]
\begin{center}
\includegraphics[width = .45\textwidth]{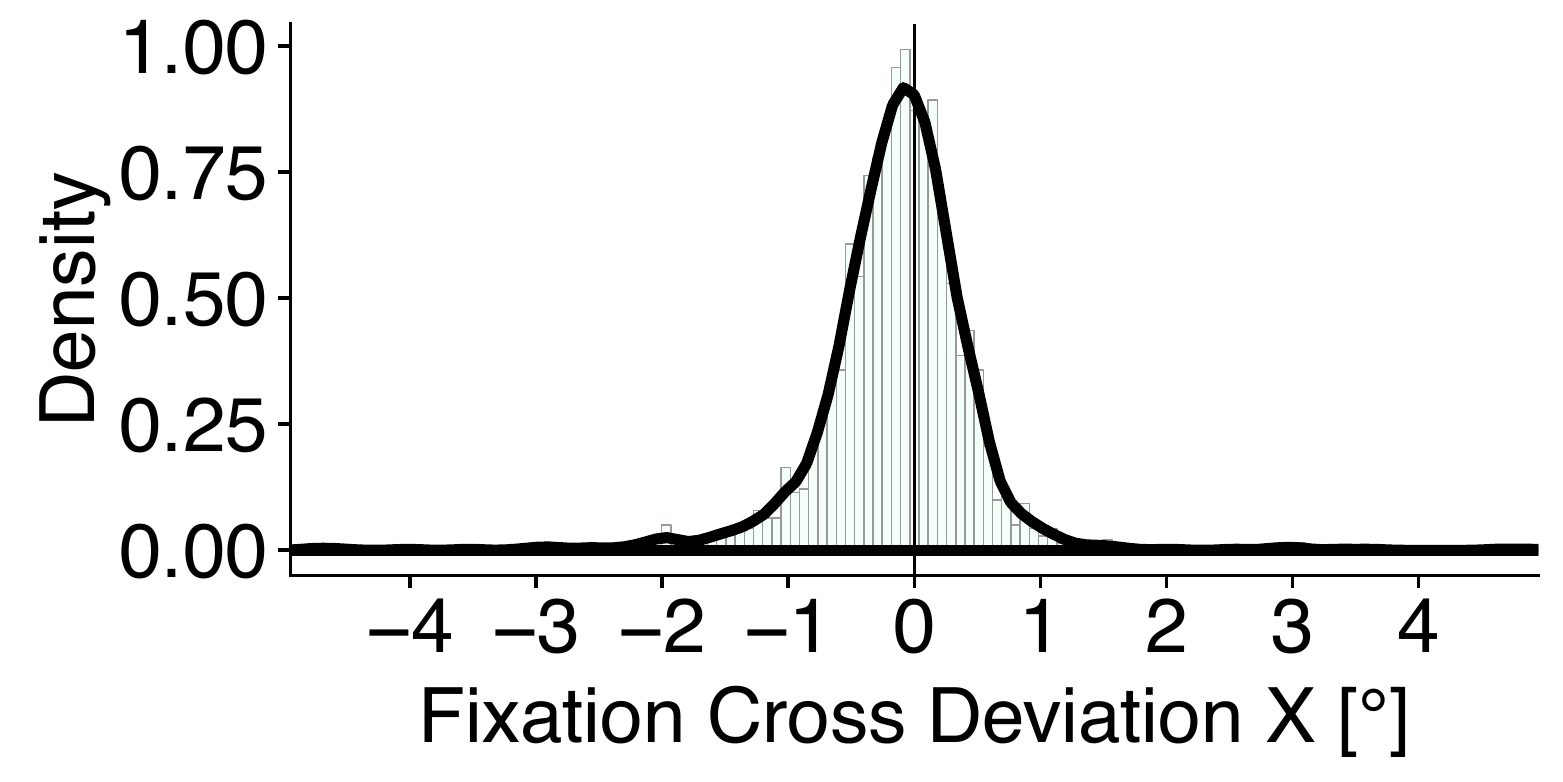}
\includegraphics[width = .45\textwidth]{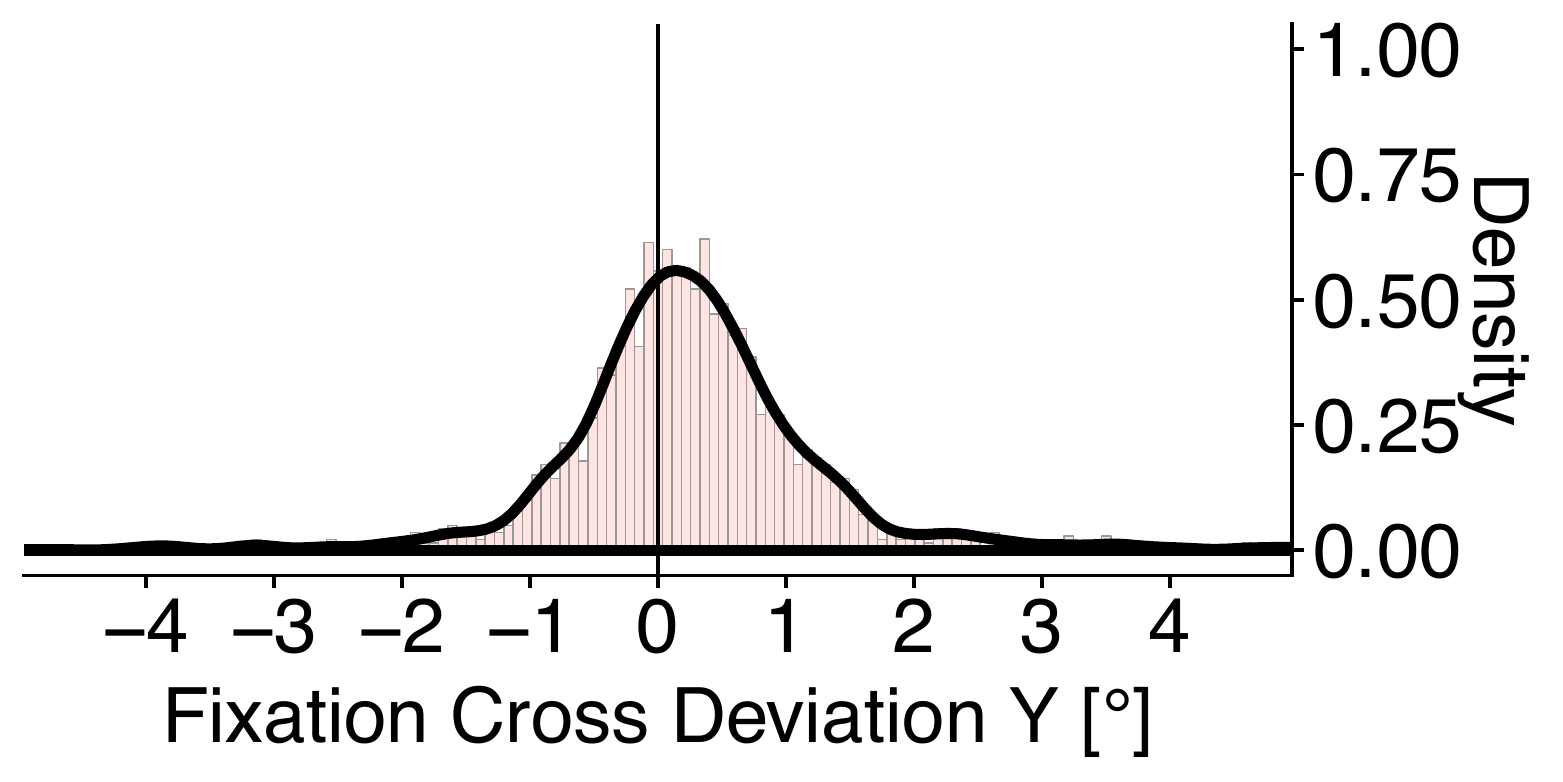}
\caption{Median horizontal and vertical deviation of participants’ gaze-position from the initial fixation cross in the left and right panel, respectively. 
\label{fig:FixCrossDeviation}}
\end{center}
\end{figure}

Finally, at least two error sources contribute to the accuracy of the measured eye-position in our setup. Measurement error generated by the eye-tracking device and the calibration procedure as well as error generated by the transformation of the eye position from scene-camera coordinates into image coordinates. To estimate the overall spatial accuracy of our setup, we calculated the deviation of participants' gaze-positions from the initial fixation cross. For each fixation check, we computed the median difference of the gaze position minus the position of the fixation cross for the last 200~ms (24 samples) of the fixation check. Figure~\ref{fig:FixCrossDeviation} shows the distributions of deviations from the initial fixation cross in the horizontal (left panel) and vertical (right panel) dimension. Horizontal deviations are mostly within 1\degree of visual angle (91.04\%) with a small leftward shift. The distribution of vertical deviations is slightly broader (76.65\% within 1\degree of visual angle) with a small upward shift. Thus, overall accuracy of our experimental setup is good but as expected somewhat weaker than in scene-viewing experiments using high-resolution eye-trackers. Note, Figure~\ref{fig:FixCrossDeviation} contains trials that were subsequently excluded from further analysis since their absolute deviation exceeded 2\degree. 

\subsection{Analyses}
Beside the analysis of fixation durations and saccade amplitudes we used three further metrics to describe the eye-movement behavior in our experiment. First, to quantify the central fixation bias \citep{tatler2007} we computed the distance to image center over time \citep{rothkegel2017}. Second, as an estimate for the overall dispersion of fixation locations on an image, we computed the informational entropy \citep{Shannon:1963}. Third, we evaluated how well fixation positions can be predicted by a distribution of fixation locations \citep{Schutt:2019}, for example, computed from a different set of fixation locations or obtained as the prediction of a computational model. \daniel{We computed linear mixed effect models (LMM) for each dependent variable using the \texttt{lme4} package \citep{lme4:2015} in \texttt{R} \citep{R:2019}. If the dependent variable deviated remarkably from a normal distribution we performed a log-transform. For the statistical model of the empirical data, we used the task as fixed factor and specified custom contrasts \citep{Schad:2018ut}. First, we compared the two Guess tasks against the two Count tasks. Second, we tested the Count Animals against the Count People condition. The third contrast coded the difference of the Guess Time and the Guess Country condition. The models were fitted by maximum likelihood estimation. For the random effect structure we ran a model selection further described in Appendix 1. Following \citep{Baayen:2008bd} we interpret all $|t|$ > 2 as significant fixed effects.}

\subsubsection{Central fixation bias}
The central fixation bias \citep{tatler2007} refers to the tendency of participants to fixate near the image center. The bias is strongest initially during a trial and reaches an asymptotic level after a few seconds. To describe this tendency we computed the mean Euclidian distance $\Delta(t)$ of the eyes to the image center over time \citep{rothkegel2017}, 
\begin{equation}
\Delta(t) = \frac{1}{m*n} \sum_{j=1}^{m}\sum_{k=1}^{n}||x_{jk}(t) - x'||
\label{eq:Delta}
\end{equation}

\noindent where $x_{jk}$ refers to the gaze coordinates of a participant $j$ on image $k$ at time $t$ and $x'$ refers to the coordinates of the image center. \daniel{If fixations were uniformly placed on an image, a value of 12~\degree~would be expected, which is the average distance of every pixel to the image center.} Note, here we chose to compute the distance to image center $\Delta(t)$ for \daniel{specific time intervals $t$: 0--400~ms, 400--800~ms, 800--1200~ms and 1200--8000~ms}. These time intervals were chosen because previous work has shown that the first 400~ms of a scanpath show more reflexive saccades in response to the image onset and after 400~ms, content or goal-driven saccades are executed \citep{rothkegel2017}. Thus these later saccades are more likely to be influenced by the specific viewing task. 

\subsubsection*{Entropy}
We use information entropy \citep{Shannon:1963} to characterize the degree of uniformity of a distribution of fixation locations. We calculate the entropy by first estimating the density of a distribution of fixation locations on a $128 \times 128$~grid. The density is computed in \texttt{R} using the \texttt{spatstat} package \citep{baddeley2005} with an optimal bandwidth for each distribution of fixation locations (\texttt{bw.scott}). After transforming the density into a probability measure (integral sums to 1), the entropy $S$ is measured in bits and computed as 
\begin{equation}
S = -\sum_{i=1}^n  {p}_i \log_2{p}_i \;,
\label{eq:entropy}
\end{equation}

\noindent where each cell $i$ of the grid is evaluated. In our analysis, an entropy of 14 bits ($n = 128 \times 128 = 2^{14}$) represents the maximum degree of uniformity, that is, the same probability of observing a fixation in each cell; a value of 0 indicates that all fixations are located in only one cell of the grid.

\subsubsection*{Predictability}
Finally, we estimated the negative cross-entropy of two fixation densities to quantify to what degree a set of fixation locations is predicted by a given probability distribution. The metric can be used to investigate how well an empirically observed fixation density (e.g., from a set of fixations recorded from other participants) or the fixation density generated by a computational model (e.g., a saliency model) predicts a set of fixation locations \citep{Schutt:2019}. The negative cross-entropy $H(p_2;p_1)$ of a set of $n$~fixations can be approximated by

\begin{equation}
H(p_2;p_1) \approx -\frac{1}{n} \sum_{i=1}^{n}\log_2\left( \hat{p}_1\big( f_2^{(i)}\big)\right) \;,
\label{eq:predict}
\end{equation}

\noindent where $\hat{p}_1$ refers to a kernel-density estimate of the fixation density $p_1$, which is evaluated at the fixation locations $f_2^{(i)}$ of a second fixation density $p_2$. The log-likelihood measure approximates how well $p_1$ approximates $p_2$ irrespective of the entropy $p_2$. We implemented the negative cross-entropy with a leave-one-subject-out cross-validation. For each participant on each image and each task we computed a separate kernel-density estimate $\hat{p}_1$ by using only the fixations of all other participants viewing the same image under the same instruction. 

In our analyses, we computed fixation densities $\hat{p}_1$ on the same $128 \times 128$ grid used for the entropy computations. All empirical densities (from sets of fixation locations) were computed in \texttt{R} using the \texttt{spatstat} package \citep{baddeley2005} with a bandwidth determined by Scott's rule for each distribution (\texttt{bw.scott}). In addition, we used fixation densities predicted by a state-of-the-art saliency model \citep{kummerer2016}. All density distributions were converted into probability distributions (intergral sums to 1) before computing the negative cross-entropy $H(p_2;p_1)$. A value of $0\frac{bit}{fix}$ demonstrates perfect predictability. A value of $-14 \frac{bit}{fix}~(128 \times 128 = 2^{14})$ is expected for a uniform probability distribution, where all locations in the probability distribution are equally likely to be fixated. In the results section we report $\Delta$ log-likelihoods that indicate the gain in predictability of the negative cross-entropy relative to a uniform distribution.

\section{Results}
In the Methods section, we ensured that the workflow necessary to measure eye movements in a relaxed version of the scene-viewing paradigm provides data quality comparable to the laboratory setup. Next, we wanted to see if it is possible to replicate task differences under this setup. As the most commonly used eye-movement parameters, we first analyzed fixation durations and saccade amplitudes. Next, we examined the distributions of fixation locations to quantify systematic differences in target selection between tasks. We compared the strength of the central fixation bias in the four tasks. A direct \daniel{within-subject} comparison of the central fixation bias on the same stimulus material has not been reported before. We computed the entropy to quantify the overall dispersion of fixation locations on an image, computed a log-likelihood to see how well fixations can be predicted across tasks and compared fixation locations in the four tasks with the predictions of a saliency model.

In our results section, \daniel{we report linear mixed effect model (LMM) analyses}. \daniel{Moreover,} we used post-hoc multiple comparisons to further investigate differences between tasks. All reported $p$-values in the multiple comparisons were adjusted according to Tukey. A summary of all investigated eye-movement parameters can be found in Table~\ref{ex:AllResults}.

\begin{table}[H]
\begin{center}
\color{black}
\caption{Mean values of eye-movement parameters under the four task instructions. The central fixation bias (CFB) is reported as the average distance $\Delta(t)$ to the image center during specific time intervals $t$.} 
\begin{tabular}{l c c c c}
& Count People  & Count Animals &   Guess Country &   Guess Time    \\
\hline
Fixation duration [ms] & 249 &  233 &  244 & 248    \\
Saccade amplitude [\degree] & 6.27 & 6.45 &   6.76 & 6.83 \\
CFB:   0--400~ms [\degree] & 5.809 &  5.573 &  5.730  & 5.596 \\
CFB: 400--800~ms [\degree] & 7.678 & 7.203 & 6.740 & 6.420 \\
CFB: 800--1200~ms [\degree] & 9.672 & 9.552 & 8.551 & 8.482  \\
CFB: 1200--8000~ms [\degree] & 10.351 & 10.899 & 9.821 & 9.688 \\
Entropy [bit] & 13.051 & 13.476 & 13.327 & 13.394  \\
Predictability [bit/fix]  & 1.187 & 0.745 & 0.936 & 0.830 \\
DeepGaze2 [bit/fix] & 0.434 & -0.101  & 0.726 & 0.562 \\
\hline
\end{tabular}
\label{ex:AllResults}
\end{center}
\end{table}

\subsection{Fixation durations}
\begin{figure}[H]
\begin{center}
\includegraphics[width = .5\textwidth]{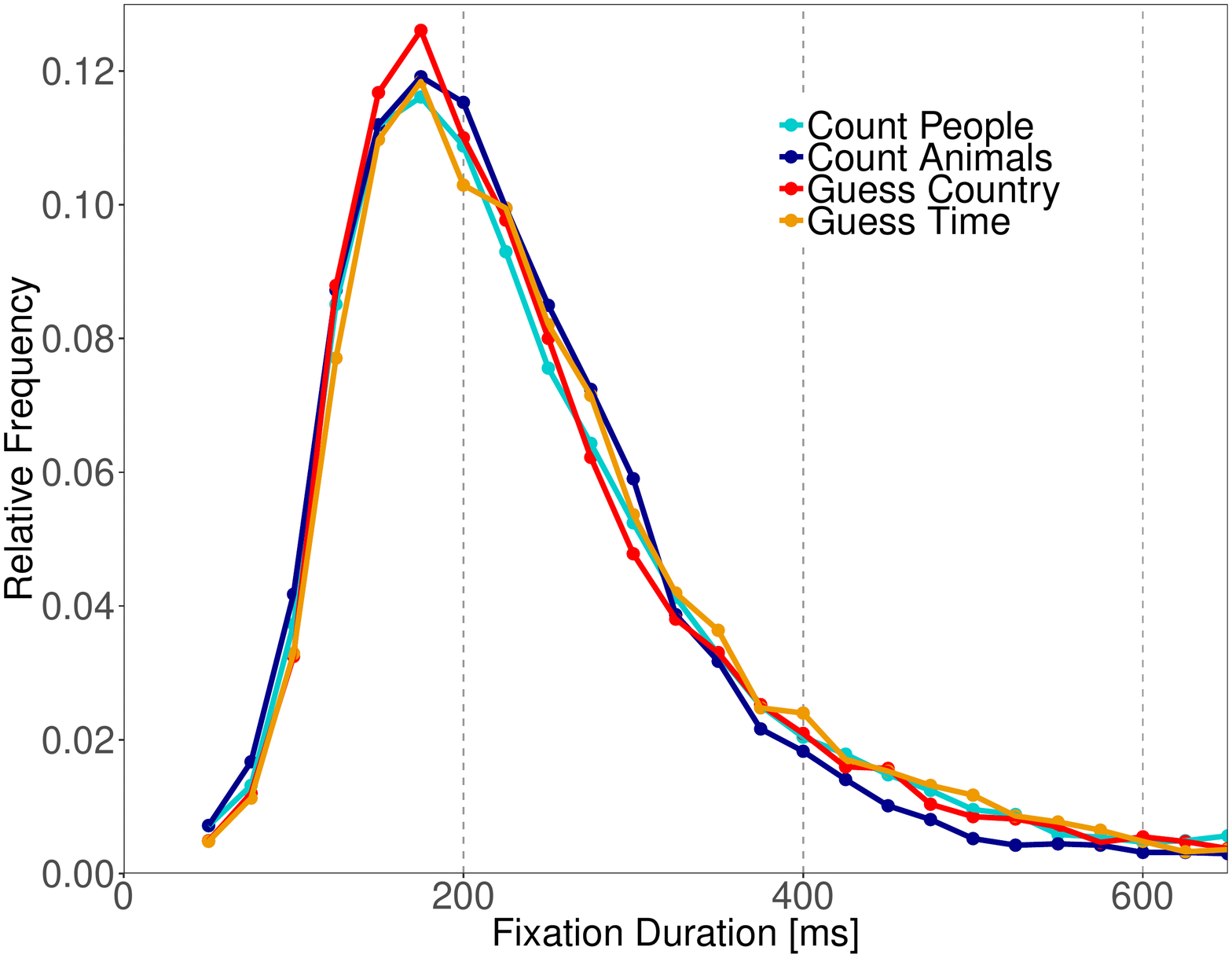}
\caption{Fixation duration distributions. The figure shows relative frequencies of fixation durations in the four tasks. Fixation durations were binned in steps of 25~ms.} 
\label{fig:FixDur}
\end{center}
\end{figure}

Distributions of fixation durations for the four different tasks are plotted in Figure~\ref{fig:FixDur}. All distributions show the characteristic form typically observed for eye movements in scene viewing. The distributions in our tasks peak at around 200~ms and show a long tail with fixation durations above 400~ms. \daniel{A linear mixed effects model (LMM) (see Methods section; \cite{lme4:2015}) revealed significant fixed effects of task (Tab.~\ref{ex:LMMFixDur}). All of our comparisons, specified by our three contrasts, show significant differences. To ensure the normal distribution of model residuals, fixation durations were log-transformed}. Fixation durations were shortest in the Count Animals condition (233~ms) and post-hoc multiple comparisons revealed that fixation durations in this task differed significantly from all other tasks \daniel{(all $p\leq 0.05$, Tab.~\ref{ex:fixdurmean})}. The effect seems to be primarily driven by a reduction of long fixation durations in the range between 350 to 550~ms (blue line in Fig.~\ref{fig:FixDur}). \daniel{There were no reliable differences in fixation durations between Count People and the Guess conditions (all $p >0.5$; Count People: 249~ms, Guess Country: 244~ms, Guess Time: 248~ms). Replicating the results from the linear mixed effect model, the Guess conditions also differed significantly in the post-hoc multiple comparisons analysis ($p <0.001$).}

\begin{table}[H]
\begin{center}
\color{black}
\caption{Fixed effects of linear mixed effect model (LMM): Fixation durations (log-transformed) for our contrasts.}
\begin{tabular}{l@{ - }l r@{.}l r@{.}l r@{.}l}
\noalign{\vskip0.3cm}
\multicolumn{2}{l}{} & \multicolumn{2}{c}{$\beta$} & \multicolumn{2}{c}{$SE$} & \multicolumn{2}{c}{$t$}\tabularnewline
\hline
Guess         & Count        &  0&02 & 0&01 &  2&16\tabularnewline
CountAnimals  & CountPeople  & -0&05 & 0&01 & -4&80\tabularnewline
GuessTime     & GuessCountry &  0&03 & 0&01 &  3&62\tabularnewline
\hline 
\end{tabular}\\
Note: $|t|$ > 2 are interpreted as significant effects.
\label{ex:LMMFixDur}
\end{center}
\end{table}

\begin{table}[H]
\begin{center}
\color{black}
\caption{Multiple comparisons of fixation durations (log-transformed) for all tasks. Adjusted p values reported (Tukey).}
\begin{tabular}{l@{ - }l r@{.}l r@{.}l r@{.}l r@{.}l}
\noalign{\vskip0.3cm}
\multicolumn{2}{l}{Posthoc comparison} &  \multicolumn{2}{c}{Estimate} & \multicolumn{2}{c}{$SE$}& \multicolumn{2}{c}{z value} &\multicolumn{2}{c}{Pr(>|z|)}    \\
\hline
Count Animals & Count People   &  -0&054  &  0&0112  &  -4&796  &   < 0&001 ***      \\
Guess Country & Count People   &  -0&017  &  0&0129  &  -1&278  &     0&564          \\
Guess Time    & Count People   &   0&010  &  0&0129  &   0&800  &     0&848          \\
Guess Country & Count Animals  &   0&037  &  0&0129  &   2&889  &     0&019 *      \\
Guess Time    & Count Animals  &   0&064  &  0&0129  &   4&983  &   < 0&001 ***   \\
Guess Time    & Guess Country  &   0&027  &  0&0074  &   3&620  &     0&002 **     \\
\hline 
\end{tabular}\\
Levels of significance:  *** <0.001, ** <0.01, * <0.05, . <0.1
\label{ex:fixdurmean}
\end{center}
\end{table}

\subsection{Saccade amplitudes}
\begin{figure}[H]
\begin{center}
\includegraphics[width = .5\textwidth]{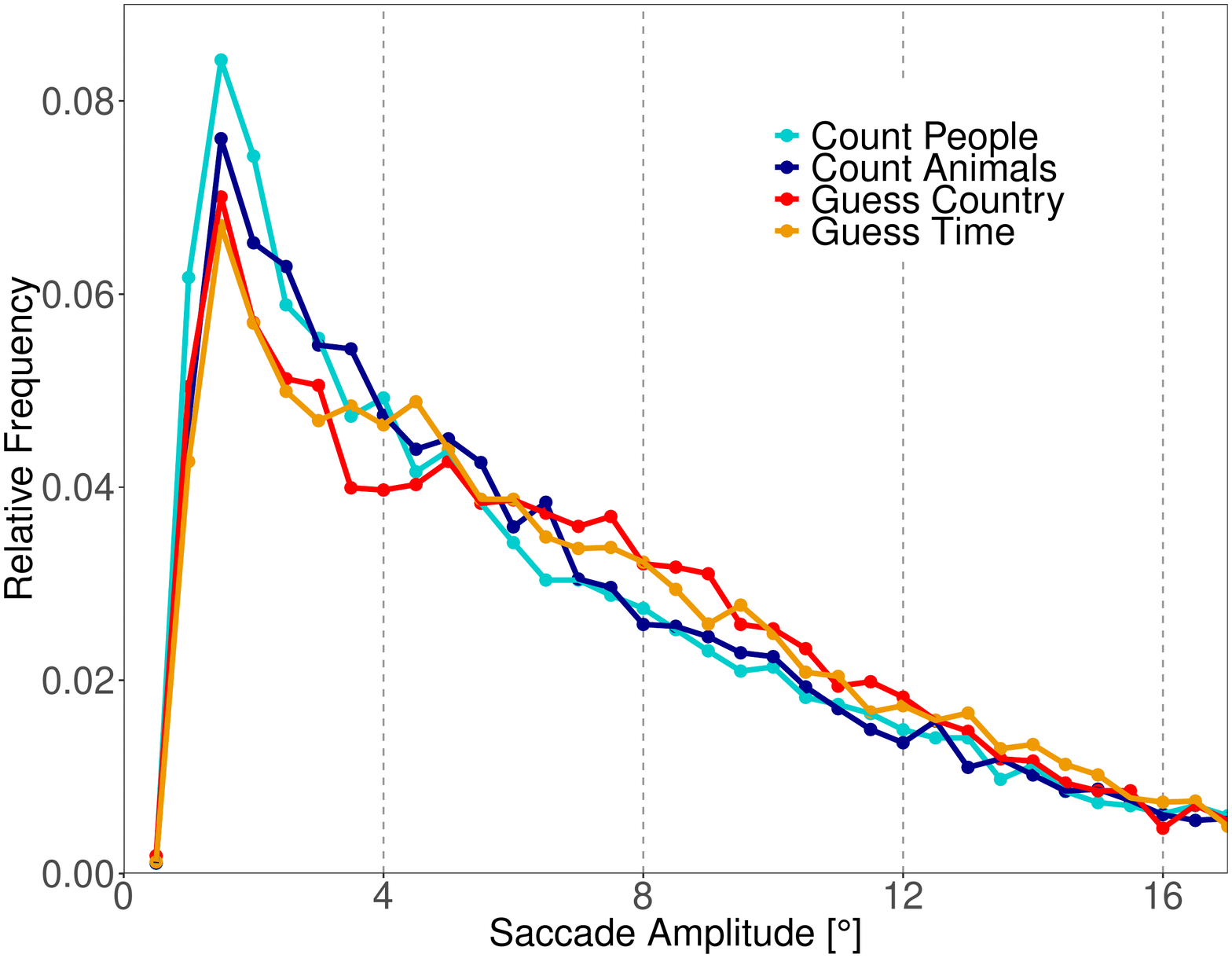}
\caption{Distribution of saccade amplitudes. The figure shows relative frequencies of saccade amplitudes in the four tasks. Saccade amplitudes were binned in steps of 0.5\degree.
\label{fig:SacAmpl}}
\end{center}
\end{figure}

Relative frequencies of saccade amplitudes for the four tasks are shown in Figure~\ref{fig:SacAmpl}. In line with previous scene-viewing experiments, saccade amplitude distributions show a peak between 2 and 3\degree~with a substantial proportion of larger saccades. \daniel{A linear mixed effect model (LMM) revealed a significant difference across the Guess and Count tasks for saccade amplitudes (log-transformed since saccade amplitudes deviated considerably from a normal distribution). Both within Guess and within Count conditions were not significant. Post-hoc multiple comparisons revealed significant differences between Count People and Guess conditions (all $p<0.001$, Tab.~\ref{ex:sacmean}). Saccade amplitudes in the Guess Country (6.76\degree) and Guess Time condition (6.83\degree) were longer on average than saccade amplitudes in the Count People (6.27\degree) condition. There were no other significant differences (all $p>0.09$).}

\begin{table}[H]
\begin{center}
\color{black}

\caption{Fixed effects of linear mixed effect model (LMM): Saccade amplitudes (log-transformed) for our contrasts.}
\begin{tabular}{l@{ - }l r@{.}l r@{.}l r@{.}l}
\noalign{\vskip0.3cm}
\multicolumn{2}{l}{} & \multicolumn{2}{c}{$\beta$} & \multicolumn{2}{c}{$SE$} & \multicolumn{2}{c}{$t$}\tabularnewline
\hline
Guess         & Count        & 0&10  &  0&03  &  3&95  \tabularnewline
CountAnimals  & CountPeople  & 0&06  &  0&04  &  1&56  \tabularnewline
GuessTime     & GuessCountry & 0&01  &  0&01  &  0&43  \tabularnewline
\hline 
\end{tabular}\\
Note: $|t|$ > 2 are interpreted as significant effects.
\label{ex:LMMSacAmpl}
\end{center}
\end{table}

\begin{table}[H]
\begin{center}
\color{black}

\caption{Multiple comparisons of saccade amplitudes (log-transformed) for all tasks. Adjusted p values reported (Tukey).}
\begin{tabular}{l@{ - }l r@{.}l r@{.}l r@{.}l r@{.}l}
\noalign{\vskip0.3cm}
\multicolumn{2}{l}{Posthoc comparison} &  \multicolumn{2}{c}{Estimate} & \multicolumn{2}{c}{$SE$}& \multicolumn{2}{c}{z value} &\multicolumn{2}{c}{Pr(>|z|)}    \\
\hline
Count Animals & Count People   &   0&059  &  0&038 &   1&560    &   0&380        \\
Guess Country & Count People   &   0&127  &  0&032 &   3&923    & < 0&001 ***    \\
Guess Time    & Count People   &   0&133  &  0&032 &   4&100    & < 0&001 ***    \\
Guess Country & Count Animals  &   0&068  &  0&032 &   2&104    &   0&138      \\
Guess Time    & Count Animals  &   0&074  &  0&032 &   2&278    &   0&093 .   \\
Guess Time    & Guess Country  &   0&006  &  0&013 &   0&430    &   0&971      \\
\hline 
\end{tabular}\\
Levels of significance:  *** <0.001, ** <0.01, * <0.05, . <0.1
\label{ex:sacmean}
\end{center}
\end{table}

\subsection{Central Fixation Bias}
\begin{figure}[H]
\begin{center}
\color{black}

\includegraphics[width = .9\textwidth]{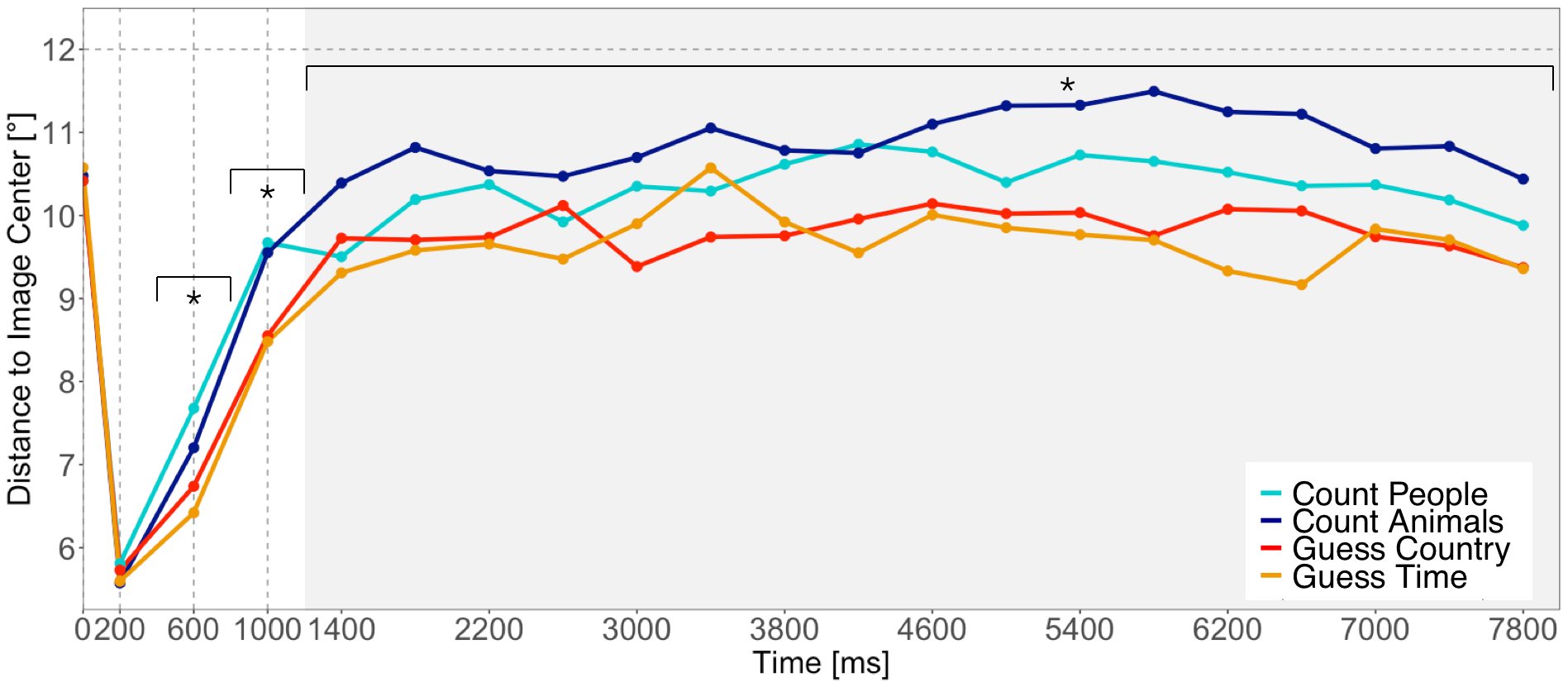}
\caption{Temporal evolution of the central fixation bias measured as the average distance to image center. Each line corresponds to one of the four instructions. The horizontal line provides the expected distance to center, if fixations were uniformly placed on an image. Level of significance: * < 0.05.
\label{fig:CFB}}
\end{center}
\end{figure}

The central fixation bias (CFB) is a systematic tendency of observers to fixate images, presented on a computer screen, near \daniel{their} center \citep{tatler2007} and is strongest during initial fixations \citep{rothkegel2017,tatler2007,tHart:2009kj}. We measured the CFB as the distance to the image center (Eq.~\ref{eq:Delta}) and found a strong initial CFB in all conditions (Fig.~\ref{fig:CFB}). Before the first saccade, participants' gaze positions were located on the initial fixation cross. The \daniel{earliest subsequent fixations} of the exploration were on average closest to the image \daniel{center}. All later fixations were less centered and the average distance to image center reached an asymptotic level after 1000 to 2000~ms. We computed the average distance of all image coordinates from the image center. A \daniel{distance to image center} of 12\degree~would be expected, if fixations were uniformly placed on the image.

We compared the distance to image center in the four tasks with \daniel{linear mixed models (LMM) for specific time intervals. There was no significant fixed effect of task during the \daniel{earliest fixations (0--400~ms, Tab.~\ref{ex:LMMCFB})}, but we observed differences between tasks for all later time intervals: for fixations in between 400--800~ms we found that Guess and Count conditions as well as Count People and Count Animals conditions differed significantly. Fixations in between 800--1200~ms differed significantly between Guess and Count conditions, but we could not find significant differences in between Guess and Count conditions. For later fixations (1200--8000~ms), all fixed effects show significant differences. Post-hoc multiple comparisons revealed no significant differences between tasks for the earliest fixations (0--400~ms) (all $p>.3$, Tab.~\ref{ex:CFB}).}

\daniel{On the following time interval (400--800~ms), fixations in Count People condition were significantly further away from the image center than fixations in both Guess conditions (all $p\leq 0.003$) and fixations in Count Animals condition were significantly further away from the image center than fixations in Guess Time condition ($p=0.003$). Additionally, in the next time interval (800--1200~ms), fixations in Count Animals condition were significantly further away from the image center than fixations in Guess Country condition ($p<.001$), but there were still no significant differences both within Guess and within Count conditions (all $p>0.8$). For the later fixations (1200-8000~ms), all tasks differed significantly (all $p\leq 0.01$).}

\begin{table}[h]
\begin{center}
\color{black}

\caption{Fixed effects of linear mixed effect models (LMM): Distance to image center across tasks for different time intervals for our contrasts.}
\begin{tabular}{l@{ - }l r@{.}l r@{.}l r@{.}l}
\multicolumn{2}{l}{Time interval} &  \multicolumn{2}{c}{$\beta$} & \multicolumn{2}{c}{$SE$}& \multicolumn{2}{c}{$t$} \\
\hline

\multicolumn{3}{l}{\textit{0--400~ms}} & \multicolumn{5}{l}{ }  \\

              Guess         & Count           &  -0&11  &  0&15  &  -0&70  \\
              CountAnimals  & CountPeople     &  -0&20  &  0&21  &  -0&96  \\
              GuessTime     & GuessCountry    &  -0&26  &  0&22  &  -1&21  \\

\multicolumn{8}{l}{} \\
\multicolumn{3}{l}{\textit{400--800~ms}} & \multicolumn{5}{l}{ }  \\

              Guess         & Count           &  -0&80  &  0&14  &   -5&92  \\
              CountAnimals  & CountPeople     &  -0&47  &  0&19  &   -2&52  \\
              GuessTime     & GuessCountry    &  -0&23  &  0&20  &   -1&15  \\

\multicolumn{8}{l}{} \\
\multicolumn{3}{l}{\textit{800--1200~ms}} & \multicolumn{5}{l}{ }  \\

              Guess         & Count          &  -1&02  &  0&18  &   -5&71  \\
              CountAnimals  & CountPeople    &  -0&01  &  0&25  &    0&05  \\
              GuessTime     & GuessCountry   &  -0&22  &  0&26  &    0&82  \\

\multicolumn{8}{l}{} \\
\multicolumn{3}{l}{\textit{1200--8000~ms}} & \multicolumn{5}{l}{ }  \\

              Guess         & Count          &  -0&89  &  0&05  &  -16&42  \\
              CountAnimals  & CountPeople    &   0&55  &  0&08  &    7&43  \\
              GuessTime     & GuessCountry   &  -0&24  &  0&08  &   -3&10  \\
\hline 
\end{tabular}\\
Note: $|t|$ > 2 are interpreted as significant effects.
\label{ex:LMMCFB}
\end{center}
\end{table}

\begin{table}[!htbp]
\begin{center}
\color{black}

\caption{Multiple comparisons of distance to image center across tasks for different time intervals. Adjusted p values reported (Tukey).}
\begin{tabular}{l@{ - }l r@{.}l r@{.}l r@{.}l r@{.}l}
\noalign{\vskip0.3cm}
\multicolumn{2}{l}{Posthoc comparison} &  \multicolumn{2}{c}{Estimate} & \multicolumn{2}{c}{$SE$}& \multicolumn{2}{c}{z value} &\multicolumn{2}{c}{Pr(>|z|)}    \\

\hline
\multicolumn{4}{l}{\textit{Fixations 0-400 ms}} & \multicolumn{6}{l}{ }  \\
Count Animals & Count People   &  -0&199  &  0&206  &  -0&963  &  0&771   \\
Guess Country & Count People   &  -0&073  &  0&214  &  -0&340  &  0&986  \\
Guess Time    & Count People   &  -0&338  &  0&211  &  -1&598  &  0&380 \\
Guess Country & Count Animals  &   0&126  &  0&214  &   0&588  &  0&936 \\
Guess Time    & Count Animals  &  -0&139  &  0&212  &  -0&655  &  0&914 \\
Guess Time    & Guess Country  &  -0&265  &  0&220  &  -1&205  &  0&623   \\
\multicolumn{10}{l}{} \\
\multicolumn{3}{l}{\textit{Fixations 400--800 ms}} & \multicolumn{6}{l}{ }  \\
Count Animals & Count People   &  -0&468  &  0&186  &  -2&518  &    0&057 .      \\
Guess Country & Count People   &  -0&923  &  0&190  &  -4&853  &  < 0&001 ***  \\
Guess Time    & Count People   &  -1&151  &  0&192  &  -6&007  &  < 0&001 ***  \\
Guess Country & Count Animals  &  -0&455  &  0&192  &  -2&370  &    0&083 .    \\
Guess Time    & Count Animals  &  -0&682  &  0&195  &  -3&507  &    0&003 **   \\
Guess Time    & Guess Country  &  -0&227  &  0&198  &  -1&147  &    0&660        \\
\multicolumn{10}{l}{} \\
\multicolumn{3}{l}{\textit{Fixations 800--1200 ms}} & \multicolumn{6}{l}{ }  \\
Count Animals & Count People   &   0&013  &  0&246  &   0&054  &    1&000       \\
Guess Country & Count People   &  -1&125  &  0&253  &  -4&452  &  < 0&001 *** \\
Guess Time    & Count People   &  -0&908  &  0&260  &  -3&491  &    0&003 ** \\
Guess Country & Count Animals  &  -1&138  &  0&248  &  -4&595  &  < 0&001 *** \\
Guess Time    & Count Animals  &  -0&922  &  0&255  &  -3&613  &    0&002 **  \\
Guess Time    & Guess Country  &   0&216  &  0&263  &   0&824  &    0&843       \\

\multicolumn{10}{l}{}\\
\multicolumn{3}{l}{\textit{Fixations 1200--8000 ms}} & \multicolumn{6}{l}{ }  \\
Count Animals & Count People   &   0&551  &  0&075  &    7&343  &  <0&001 ***  \\
Guess Country & Count People   &  -0&496  &  0&078  &   -6&345  &  <0&001 ***  \\
Guess Time    & Count People   &  -0&741  &  0&077  &   -9&588  &  <0&001 ***   \\
Guess Country & Count Animals  &  -1&048  &  0&077  &  -13&672  &  <0&001 ***   \\
Guess Time    & Count Animals  &  -1&293  &  0&076  &  -17&029  &  <0&001 ***   \\
Guess Time    & Guess Country  &  -0&245  &  0&079  &   -3&104  &   0&010 *     \\
\hline

\end{tabular}\\
Levels of significance:  *** <0.001, ** <0.01, * <0.05, . <0.1
\label{ex:CFB}
\end{center}
\end{table}

\subsection{Entropy}
\begin{figure}[H]
\begin{center}
\includegraphics[width = .7\textwidth]{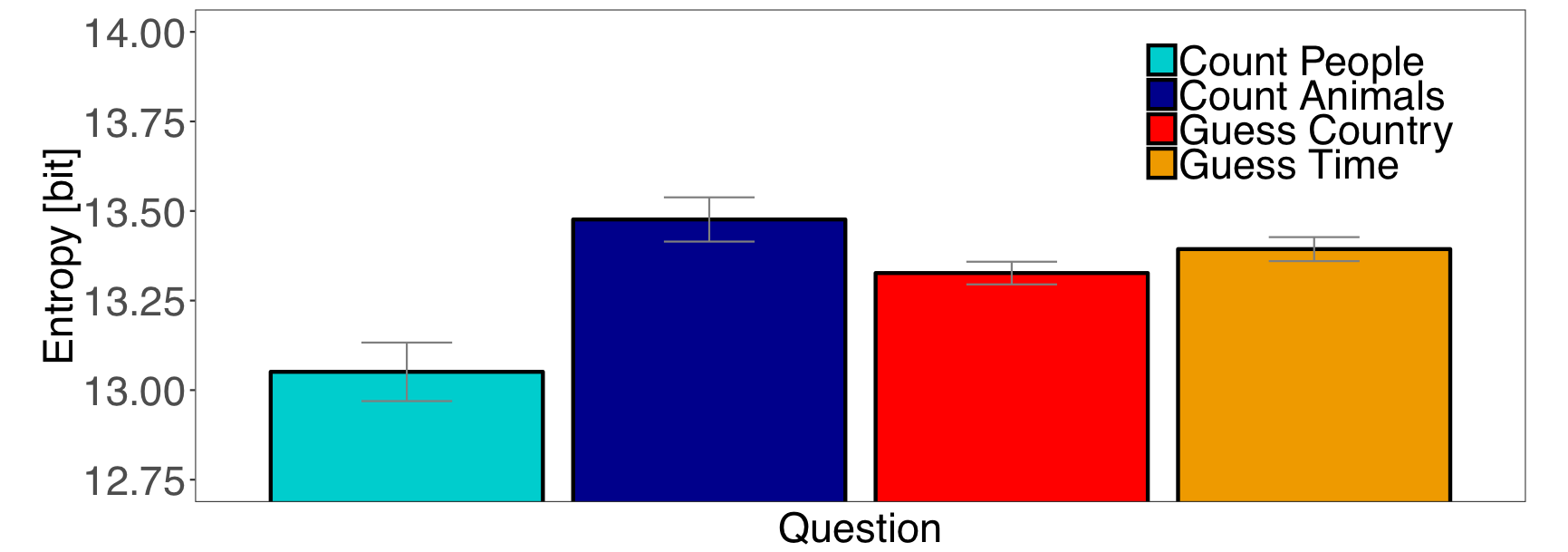}
\caption{Shannon's entropy. Average entropy of fixation densities on an image in the four tasks. A value of 14 bit is expected for a uniform fixation density. Smaller values indicate that fixations cluster in specific parts of an image. Confidence intervals were corrected for within-subject designs \citep{Cousineau:2005,Morey:2008}. 
\label{fig:Entropy}}
\end{center}
\end{figure}

We computed Shannon's entropy, Eq.~(\ref{eq:entropy}), as a measure to describe the overall distribution of fixation locations on an image (Fig.~\ref{fig:Entropy}). If all fixations are at the same location, Shannon's entropy would be 0~bit. If all locations are fixated equally often, i.e., distributed uniformly, a value of 14~bit would be expected. The entropy of fixations locations in the Count People condition differed the most from a uniform distribution (13.051~bit). The entropy of the Count Animals condition was closest to a uniform distribution (13.476~bit). The values of the entropy of Guess Country (13.327~bit) and Guess Time (13.394~bit) lay between the two Count tasks. \daniel{A linear mixed effect model (LMM) comparing the entropy of the four tasks showed significant differences across all our contrasts. Fixations in Guess conditions are significantly more distributed over the images than fixations in Count conditions ($t=2.12$, Tab.~\ref{ex:LMMEntropy}). Fixations in Count Animals condition are more widely spread over the images than those from Count People condition ($t=3.73$) and fixations in Guess Country task are more distributed than fixation locations measured in Guess Time task ($t=2.06$).} Post-hoc multiple comparison analysis (Tab.~\ref{ex:entropy}) revealed that the Count People condition differed significantly from all other conditions (all $p\leq 0.001$). There were no other significant differences between tasks (all $p>.1$).

\begin{table}[H]
\begin{center}
\color{black}

\caption{Fixed effects of linear mixed effect model (LMM): Entropy for our contrasts.}
\begin{tabular}{l@{ - }l r@{.}l r@{.}l r@{.}l}
\noalign{\vskip0.3cm}
\multicolumn{2}{l}{} & \multicolumn{2}{c}{$\beta$} & \multicolumn{2}{c}{$SE$} & \multicolumn{2}{c}{$t$}\tabularnewline
\hline
Guess         & Count        & 0&09  &  0&04  &  2&12  \tabularnewline
CountAnimals  & CountPeople  & 0&39  &  0&11  &  3&73  \tabularnewline
GuessTime     & GuessCountry & 0&07  &  0&03  &  2&06  \tabularnewline
\hline 
\end{tabular}\\
Note: $|t|$ > 2 are interpreted as significant effects.
\label{ex:LMMEntropy}
\end{center}
\end{table}

\begin{table}[H]
\begin{center}
\color{black}

\caption{Multiple comparisons of entropy for all tasks. Adjusted p values reported (Tukey).}
\begin{tabular}{l@{ - }l r@{.}l r@{.}l r@{.}l r@{.}l}
\noalign{\vskip0.3cm}
\multicolumn{2}{l}{Posthoc comparison} &  \multicolumn{2}{c}{Estimate} & \multicolumn{2}{c}{$SE$}& \multicolumn{2}{c}{z value} &\multicolumn{2}{c}{Pr(>|z|)}    \\
\hline
Count Animals & Count People   &   0&394  &  0&105  &   3&733 & < 0&001 ***    \\
Guess Country & Count People   &   0&255  &  0&071  &   3&612 &   0&001 **     \\
Guess Time    & Count People   &   0&324  &  0&071  &   4&599 & < 0&001 ***    \\
Guess Country & Count Animals  &  -0&139  &  0&071  &  -1&967 &   0&178      \\
Guess Time    & Count Animals  &  -0&069  &  0&071  &  -0&981 &   0&736     \\
Guess Time    & Guess Country  &   0&070  &  0&034  &   2&060 &   0&147      \\
\hline 
\end{tabular}\\
Levels of significance:  *** <0.001, ** <0.01, * <0.05, . <0.1
\label{ex:entropy}
\end{center}
\end{table}

\subsection{Predictability}
\begin{figure}[H]
\begin{center}
\includegraphics[width = .9\textwidth]{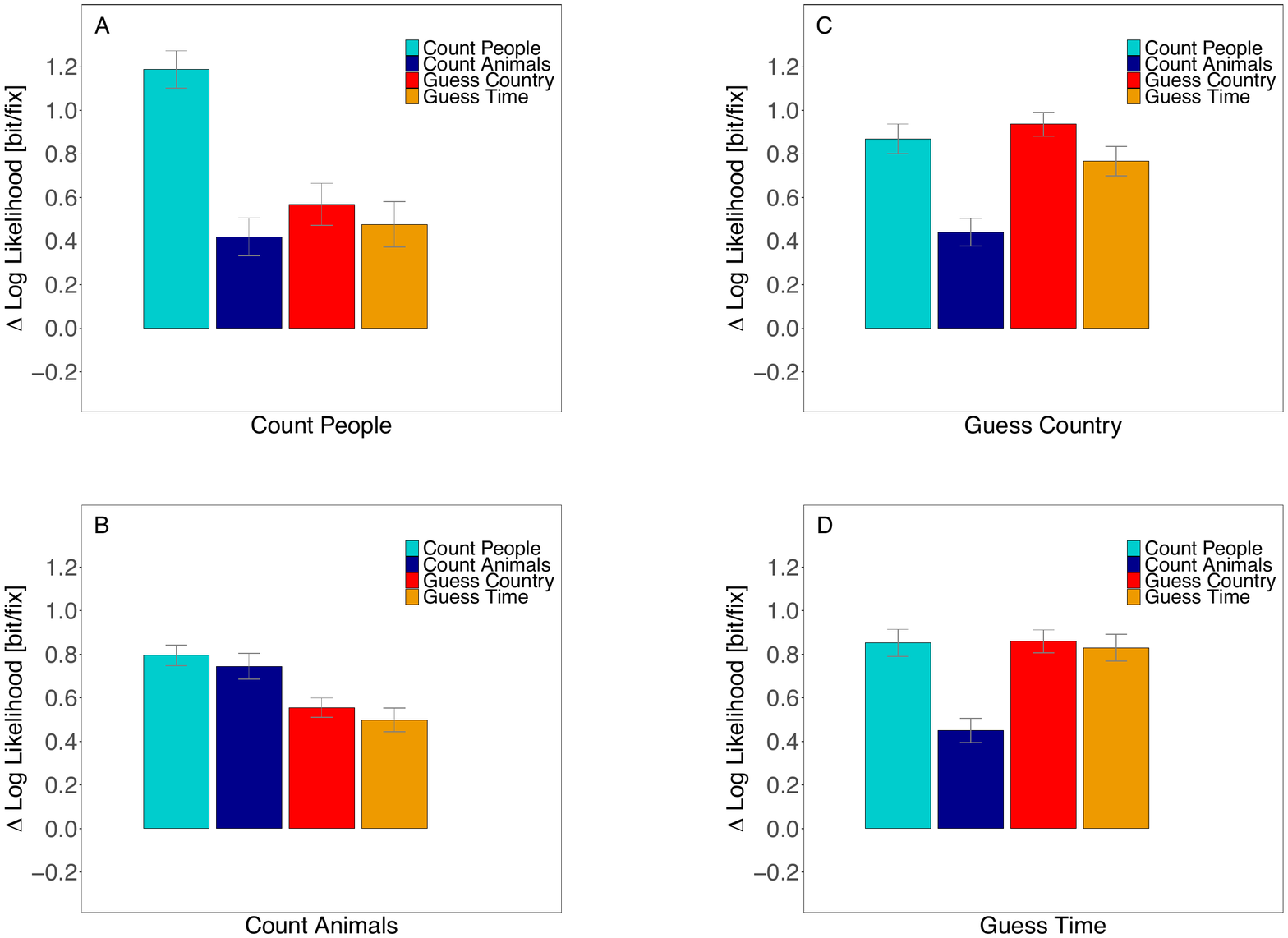}
\caption{Average predictability of fixation locations in a task. Predictability was measured in bit per fixation as the average gain in log-likelihood of each fixation relative to a uniform distribution. Fixations were predicted from the distribution of all fixation locations measured under (A) the Count People, (B) Count Animals, (C) Guess Country, and (D) Guess Time instruction. Confidence intervals were corrected for within-subject designs \citep{Cousineau:2005,Morey:2008}.
\label{fig:PredictPerTask}}
\end{center}
\end{figure}

Next, we computed negative cross-entropies of fixation densities to investigate how well fixation locations from one observer viewing an image under a specific instruction can be predicted by the distribution of fixation locations from other observers viewing the same image under one of the four instructions (Fig.~\ref{fig:PredictPerTask}). Panels correspond to how well fixation locations are predicted by the distribution of all other observers viewing an image under the Count People (panel A), Count Animals (B), Guess Country (C), and Guess Time instruction (D). We report log-likelihood differences, which give the average gain in the log-likelihood per fixation relative to a uniform distribution (Eq.~\ref{eq:predict}).

In a first step, we compared how well fixations of one observer viewing an image were on average predicted by other observers viewing the same image under the same instruction. The values correspond to the cyan bar in panel A, the blue bar in panel B, the red bar in panel C, and the orange bar in panel D (Fig.~\ref{fig:PredictPerTask}). \daniel{A linear mixed effect model revealed that both within Guess and within Count conditions differ significantly from each other. Fixations in Count People condition are better predictable than those in Count Animals condition ($t=-4.54$, Tab.~\ref{ex:LMMPredict}) And fixations in Guess Country condition are better to predict than fixations of Guess Time task ($t=-2.24$)}
Post-hoc multiple comparisons (Tab.~\ref{ex:predictall}) revealed that predictability of fixation locations differed significantly between all tasks (all $p\leq 0.025$) except for the two Guess conditions ($p=0.104$) and the Guess Time and Count Animals condition comparison ($p=0.209$). Thus, when fixations were predicted by other observers viewing an image under the same instruction, fixations from the Count People condition ($1.19~\frac{bit}{fix}$) were better predicted than fixations in the Guess Country ($0.94~\frac{bit}{fix}$) and Guess Time ($0.83~\frac{bit}{fix}$) condition, which in turn were better predicted than fixations in the Count Animals condition ($0.74~\frac{bit}{fix}$).

\begin{table}[H]
\begin{center}
\color{black}

\caption{Fixed effects of linear mixed effect model (LMM): Predictability for our contrasts.}
\begin{tabular}{l@{ - }l r@{.}l r@{.}l r@{.}l}
\noalign{\vskip0.3cm}
\multicolumn{2}{l}{} & \multicolumn{2}{c}{$\beta$} & \multicolumn{2}{c}{$SE$} & \multicolumn{2}{c}{$t$}\tabularnewline
\hline
Guess         & Count        & -0&03  &  0&05  &  -0&62 \tabularnewline
CountAnimals  & CountPeople  & -0&39  &  0&09  &  -4&54 \tabularnewline
GuessTime     & GuessCountry & -0&08  &  0&04  &  -2&24 \tabularnewline
\hline 
\end{tabular}\\
Note: $|t|$ > 2 are interpreted as significant effects.
\label{ex:LMMPredict}
\end{center}
\end{table}

\begin{table}[H]
\begin{center}
\color{black}

\caption{Multiple comparisons of predictability for all tasks. Adjusted p values reported (Tukey).}
\begin{tabular}{l@{ - }l r@{.}l r@{.}l r@{.}l r@{.}l}
\noalign{\vskip0.3cm}
\multicolumn{2}{l}{Posthoc comparison} &  \multicolumn{2}{c}{Estimate} & \multicolumn{2}{c}{$SE$}& \multicolumn{2}{c}{z value} &\multicolumn{2}{c}{Pr(>|z|)}    \\
\hline
Count Animals & Count People   &  -0&388   &    0&086  &  -4&538  &  < 0&001 ***   \\
Guess Country & Count People   &  -0&182   &    0&066  &  -2&782  &    0&025 *     \\
Guess Time    & Count People   &  -0&263   &    0&066  &  -4&015  &  < 0&001 ***   \\
Guess Country & Count Animals  &   0&206   &    0&065  &   3&147  &    0&008 **    \\
Guess Time    & Count Animals  &   0&125   &    0&065  &   1&913  &    0&209      \\
Guess Time    & Guess Country  &  -0&081   &    0&036  &  -2&240  &    0&104       \\

\hline 
\end{tabular}\\
Levels of significance:  *** <0.001, ** <0.01, * <0.05, . <0.1
\label{ex:predictall}
\end{center}
\end{table}

In a second step, we investigated whether predictions of the same task differed from the predictions of other tasks. Figure~\ref{fig:PredictPerTask}A shows how well the distribution of fixation locations from the Count People condition predicted fixation locations of another observer viewing the same image under one of the four instructions. As expected, the distribution of fixation locations from the Count People condition predicted fixation locations in the Count People condition better than fixations in any other condition ($\sim 1.2~\frac{bit}{fix}$ vs. $\sim 0.5~\frac{bit}{fix}$). We computed \daniel{a linear mixed effect model (LMM)} with treatment contrasts \daniel{of the fixed factors} to test the deviations from the Count People condition. Our analysis confirmed that all conditions differed significantly from the Count People condition \daniel{(all $|t|\geq 20.92$, Tab.~\ref{ex:predict})}.

\begin{table}[H]
\begin{center}
\color{black}

\caption{Fixed effects of linear mixed effect models (LMM): Predictability with treatment contrasts for the gain in log-likelihood over a uniform distribution. Each block represents the predictions based on the distribution of fixation locations from one task. The intercept corresponds to a prediction of the same task, treatment contrasts represent deviations from this prediction.}
\begin{tabular}{l@{ - }l r@{.}l r@{.}l r@{.}l}
\multicolumn{2}{l}{Treatment comparison} &  \multicolumn{2}{c}{$\beta$} & \multicolumn{2}{c}{$SE$}& \multicolumn{2}{c}{$t$} \\
\hline

\multicolumn{3}{l}{\textit{Count People}} & \multicolumn{5}{l}{ }  \\

\multicolumn{2}{l}{Intercept: Count People}  &   1&19  &  0&07  &   16&89  \\
              Count Animals & Count People   &  -0&77  &  0&03  &  -28&17  \\
              Guess Country & Count People   &  -0&59  &  0&03  &  -20&92  \\
              Guess Time    & Count People   &  -0&69  &  0&03  &  -24&57  \\

\multicolumn{8}{l}{} \\
\multicolumn{3}{l}{\textit{Count Animals}} & \multicolumn{5}{l}{ }  \\

\multicolumn{2}{l}{Intercept: Count Animals}  &   0&75  &  0&06  &   12&74  \\
              Count People  & Count Animals   &   0&06  &  0&02  &    3&81  \\
              Guess Country & Count Animals   &  -0&19  &  0&02  &  -11&83  \\
              Guess Time    & Count Animals   &  -0&24  &  0&02  &  -14&97  \\

\multicolumn{8}{l}{} \\
\multicolumn{3}{l}{\textit{Guess Country}} & \multicolumn{5}{l}{ }  \\

\multicolumn{2}{l}{Intercept: Guess Country}  &   0&94  &  0&07  &   13&48  \\
              Count People  & Guess Country   &  -0&05  &  0&02  &   -2&70  \\
              Count Animals & Guess Country   &  -0&50  &  0&02  &  -26&16  \\
              Guess Time    & Guess Country   &  -0&14  &  0&02  &   -7&26  \\

\multicolumn{8}{l}{} \\
\multicolumn{3}{l}{\textit{Guess Time}} & \multicolumn{5}{l}{ }  \\

\multicolumn{2}{l}{Intercept: Guess Time}  &   0&86  &  0&06  &   13&53  \\
              Count People  & Guess Time   &   0&02  &  0&02  &    1&02  \\
              Count Animals & Guess Time   &  -0&40  &  0&02  &  -22&57  \\
              Guess Country & Guess Time   &   0&01  &  0&02  &    0&77  \\
\hline 
\end{tabular}\\
Note: $|t|$ > 2 are interpreted as significant effects.
\label{ex:predict}
\end{center}
\end{table}

\daniel{Likewise, our analysis confirmed that all conditions differed significantly from the Guess Country condition. Figure~\ref{fig:PredictPerTask}C shows that the distribution of fixation locations from the Guess Country condition differ significantly in their prediction of fixation locations of another observer viewing the same image under the Guess Country task versus one of the other instructions (all $|t|\geq 2.70$). While fixation locations were best predicted by the same task in the Count People and the Guess Country conditions, the results for the other conditions were less clear-cut. For Count Animals condition (Fig.~\ref{fig:PredictPerTask}B) we also found significance across all treatment contrasts (all $|t|\geq 3.81$), but the distribution of fixation locations from the Count Animals condition predicted fixation locations of other observers viewing the same image under Count People condition better than fixation locations of other observer viewing the same image under Count Animals condition ($\beta = 0.06$). And finally, predictions of the Guess Time condition (Fig.~\ref{fig:PredictPerTask}D) did not reveal differences between Guess Time, Guess Country and Count People (all $|t|\leq~1.02$), while predictions of fixation locations in the Count Animals condition were significantly reduced ($t=-22.57$). }

\subsection{Saliency}
\begin{figure}[H]
\begin{center}
\includegraphics[width = .5\textwidth]{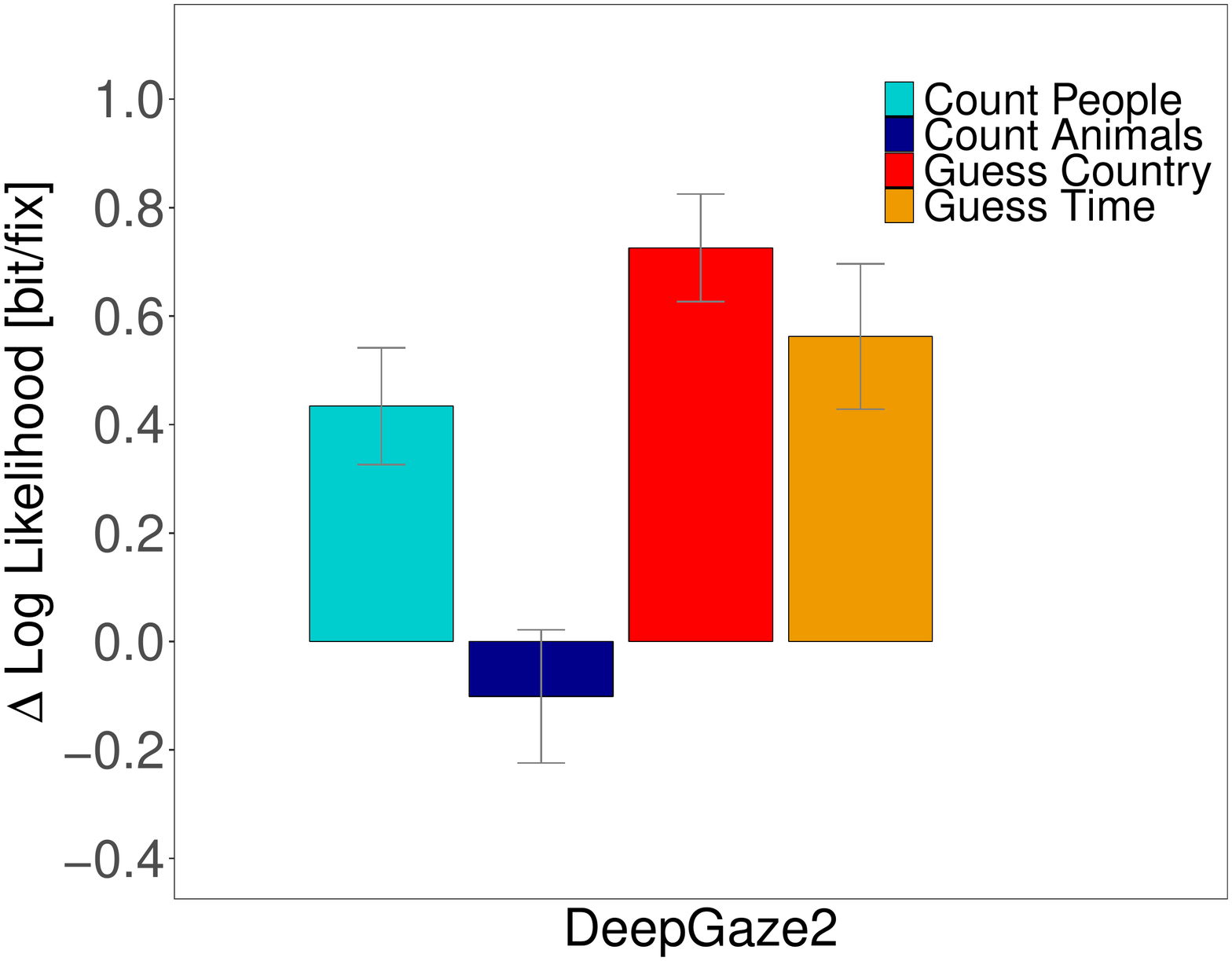}
\caption{Average predictability of fixation locations in each task by the DeepGaze2 model. Predictability was measured in bit per fixation as the average gain in log-likelihood of each fixation relative to a uniform distribution.  Confidence intervals were corrected for within-subject designs \citep{Cousineau:2005,Morey:2008}.\label{fig:DeepGaze}}
\end{center}
\end{figure}

Finally, we evaluated whether fixation locations in the four tasks can be predicted by the currently most successful saliency model \citep[DeepGaze2,][]{kummerer2016}. For each task, we computed the log-likelihood gain of the DeepGaze2 model over a uniform prediction (Fig.~\ref{fig:DeepGaze}). We choose DeepGaze2 on the basis that it is currently the best-performing saliency model in the MIT-saliency benchmark \citep{mit-saliency-benchmark} \daniel{and selected the model option that took the central fixation bias from the MIT1003 dataset \citep{Judd.CompVision.2009} into account.} Images were downsampled to $128 \times 128$ pixels, uploaded to the authors' web interface \url{deepgaze.bethgelab.org} that provided the model predictions.  As the predictions are computed in units of natural logarithm, we converted all log-likelihoods to base  2.  

\begin{table}[H]
\begin{center}
\color{black}

\caption{Fixed effects of linear mixed effect model (LMM): DeepGaze2 predictability gain for our contrasts.}
\begin{tabular}{l@{ - }l r@{.}l r@{.}l r@{.}l}
\noalign{\vskip0.3cm}
\multicolumn{2}{l}{} & \multicolumn{2}{c}{$\beta$} & \multicolumn{2}{c}{$SE$} & \multicolumn{2}{c}{$t$}\tabularnewline
\hline
Guess         & Count        &  0&46  &  0&08  &  6&11  \tabularnewline
CountAnimals  & CountPeople  & -0&59  &  0&15  & -4&07  \tabularnewline
GuessTime     & GuessCountry & -0&19  &  0&07  & -2&61  \tabularnewline
\hline 
\end{tabular}\\
Note: $|t|$ > 2 are interpreted as significant effects.
\label{ex:LMMDeepGaze}
\end{center}
\end{table}

Since DeepGaze2 was developed to predict eye movements in scene viewing, our results show that fixation locations in the Guess Country condition were most similar to fixation locations in scene viewing ($\sim 0.7~\frac{bit}{fix}$). Fixation locations in the Guess Time and Count People condition were also predicted better than by a uniform distribution ($\sim 0.5~\frac{bit}{fix}$ \& $\sim 0.4~\frac{bit}{fix}$). In contrast, fixation locations in the Count Animals condition were not well predicted by DeepGaze2. Performance was not better than predictions by a uniform distribution of fixation locations ($\sim -0.1~\frac{bit}{fix}$). \daniel{A linear mixed effect model revealed significant differences of our three specified contrasts. Fixation locations in Guess conditions can be better predicted by DeepGaze2 than in Count conditions ($t=6.11$, Tab.~\ref{ex:LMMDeepGaze}). Predictions of fixation locations in Count People task differ significantly from Count Animals task ($t=-4.07$) and fixation locations of Guess Country condition showed better predictability by DeepGaze2 than fixation locations of Guess Time conditions ($t=-2.16$).} Post-hoc multiple comparisons are listed in Table~\ref{ex:DeepGaze}. Predictability of fixation locations differed significantly between all tasks (all $p<0.05$) except for the Count People \daniel{and the Guess conditions (all $p>0.08$).}

\begin{table}[H]
\begin{center}
\color{black}

\caption{Multiple comparisons of DeepGaze2 predictability gain for all tasks. A djusted p values reported (Tukey).}
\begin{tabular}{l@{ - }l r@{.}l r@{.}l r@{.}l r@{.}l}
\noalign{\vskip0.3cm}
\multicolumn{2}{l}{Posthoc comparison} &  \multicolumn{2}{c}{Estimate} & \multicolumn{2}{c}{$SE$}& \multicolumn{2}{c}{z value} &\multicolumn{2}{c}{Pr(>|z|)}    \\
\hline
Count Animals & Count People   &  -0&591  &  0&145  &  -4&066  &  <0&001 ***    \\
Guess Country & Count People   &   0&261  &  0&111  &   2&350  &   0&082 .      \\
Guess Time    & Count People   &   0&074  &  0&111  &   0&666  &   0&906        \\
Guess Country & Count Animals  &   0&852  &  0&111  &   7&678  &  <0&001 ***    \\
Guess Time    & Count Animals  &   0&665  &  0&111  &   5&996  &  <0&001 ***   \\
Guess Time    & Guess Country  &  -0&187  &  0&072  &  -2&609  &   0&041 *      \\
\hline 
\end{tabular}\\
Levels of significance:  *** <0.001, ** <0.01, * <0.05, . <0.1
\label{ex:DeepGaze}
\end{center}
\end{table}

\section{Discussion}
Eye movements during scene viewing are typically studied to investigate the allocation of visual attention on natural, ecologically valid stimuli while keeping the benefits of a highly controlled laboratory setup. However, several aspects of the scene-viewing paradigm have been criticized that question the generalizability of results and a paradigmatic shift towards the study of natural tasks has been proposed \citep{tatler2011}. Here, we demonstrate how to adapt the scene-viewing paradigm to make a \daniel{smooth} transition from the scene-viewing paradigm to more natural tasks. This transition allows to keep the high experimental control of a laboratory setting, bases new research on a solid theoretical ground and simultaneously deals with the limitations of the classical scene-viewing paradigm. 
 
As a starting point, we demonstrated the general viability of our approach, where we used mobile eye-tracking and a projective transformation to convert gaze coordinates from head-centered coordinates into image-centered coordinates. In the experiment, participants were allowed to move their body and head, since we took away the chin rest, but we did not induce interaction with the stimulus material, which might have produced different gaze patterns \citep{Epelboim:1995vc}. \rev{In the presence of such interaction, the control of the gaze deployment system might be rather different. Therefore, we kept interaction at minimum in the current study. However, care has to be taken in follow-up studies that include forms of interaction with stimuli for even more natural behavior.} They viewed the same images under four different instructions. We implement two counting instructions, where \daniel{participants} had to determine the number of people or animals present in a given image. In the two remaining conditions, participants were asked to guess the country, where the given image was taken, or the time of day, at which the image was recorded. Our analyses replicated the sensitivity of various eye-movement measures to specific tasks \citep{castelhano2009,DeAngelus:2009bm,Mills:2011}. We observed differences between tasks in fixation durations, saccade amplitudes, strength of the central fixation bias, and in eye-movement measures related to distributions of fixation locations. Furthermore, fixation locations in the four tasks were reasonably well predicted by a recent saliency model \citep{kummerer2016}. 

\subsection{Central fixation bias across tasks}
An important observation in our study concerned the central fixation bias \citep{tatler2007}. While it is well documented that viewers prefer to fixate near the center of images and that this behavior generalizes to other tasks \citep{Ioannidou.JEMR.2016}, a direct \daniel{within-subject} comparison of the central fixation bias across tasks on the same stimulus material has not been reported before. As the central fixation bias typically is strongest during initial fixations \citep{rothkegel2017,tatler2007,tHart:2009kj}, we investigated the temporal evolution of the central fixation bias in the four tasks. We observed a strong initial response towards the image center on the \daniel{earliest fixations} and found no differences in the strength of the early central fixation bias between tasks. The central fixation bias decreased on \daniel{later} fixations and reached an asymptotic behavior after \daniel{1000--2000~ms}. Interestingly, from the second \daniel{inspected time interval (400--800~ms)} onwards the central fixation bias depended on the task given to a participant. Our data suggest a task-independent early central fixation bias and a later task-dependent central fixation bias that reflects differences in the selection of fixation locations during exploration. 

\subsection{Predictability of fixation locations across tasks} 
Since their seminal work \citep{buswell1935,yarbus1967eye} it has been known that eye movements on an image depend on the instruction given to an observer. While task differences have often been replicated
\citep{castelhano2009,DeAngelus:2009bm,Mills:2011}, prediction of a specific task from a given eye-movement trace \daniel{has resulted in incoherent success}. \rev{While \citet{greene2012} reported a failure to recover task from eye movements reliably, \citet{Borji.JVis.2014} demonstrated successful prediction of task from eye movements using the same data set.} Here, we investigated how well fixation locations can be predicted by the distribution of fixation locations from other participants viewing an image under the same or a different instruction \citep{Schutt:2019}. We made three important observations. 

First, when fixation locations were predicted by fixations of other observers viewing an image under the same instruction, predictability of fixation locations differed across tasks. The log-likelihood gain relative to a uniform distribution was highest in the Count People condition, lowest in the Count Animals condition, and in between in the two Guess conditions. Thus, there was no simple relation in predictability between the Count and Guess instructions. The entropy of the fixation location distributions resembled this result. Fixation locations deviated the most from a uniform distribution in the Count People condition and deviated the least from a uniform distribution in the Count Animals condition. Thus, predictability in our tasks can at least partially be explained by the degree  of aggregation of fixation locations in the four tasks. It is important to note, however, that this relation is not mandatory, as the entropy only affects the upper limit of the predictability measure. Our results demonstrate that the chosen task influences the inter-observer predictability of fixation locations and confirms the need to deliberately choose an instruction in the scene-viewing paradigm that is appropriate for the research question. 

Second, we compared predictability of fixation locations across tasks. In general, log-likelihood gains were highest for fixation locations predicted by other participants viewing an image under the same instruction \daniel{in the majority of tasks}. However, fixation location distributions from \daniel{half of the} tasks were not very specific in their predictions and log-likelihood gains for at least one other task were as high as the log-likelihood gains for the task itself \daniel{or another log-likelihood gain for another task was higher.} Thus, while it is possible to find tasks that lead to very different distributions of fixation locations \citep{buswell1935,yarbus1967eye}, many tasks will result in overlapping distributions, \daniel{at least on static images in a laboratory setup}. The strong overlap in fixation locations between some tasks makes it difficult to differentiate these tasks on the basis of their fixation locations.

Third, fixation locations recorded in the Count People condition showed a \daniel{distinct} pattern. While fixation locations from the Count People condition were well predicted by all other tasks, fixations from the Count People condition primarily predicted fixations from the task itself. We believe that this asymmetry arose from the peculiar role of people and faces for eye movements on images. It is well known that people and faces attract gaze in scene viewing \citep{Cerf.AdvNeuralInfoProcSyst.2007,Judd.CompVision.2009} and that at least some of these fixations are placed involuntarily \citep{Cerf.JVis.2009}. \daniel{ \cite{torralba2006} showed that participants that had to count the number of people in a scene used their prior spatial knowledge and directed their fixations toward locations likely to contain people.} \rev{As a consequence, increased fixation probabilities might be caused by expectations of faces/people rather than the actual existence of corresponding features. This effect might even be enhanced in the Count People task, which puts a particular emphasis on people and locations with high expectations to find people; so it is likely that participants made even more fixations in related regions.} This interpretation is supported by the low entropy in the Count People condition, which indicates that fixations clustered more in the Count People task than in any other task. Since people and faces attracted gaze in all tasks and in particular in the Count People condition, all tasks were well able to predict fixation locations in the Count People condition. At the same time, the Count People condition mostly predicted fixations on people and faces in the other conditions. Since these  are only a fraction of all fixations in the other conditions, predictability performance of the Count People condition was relatively low for these tasks.

\subsection{Search vs.~free viewing}
Images in our experiment were viewed under four different instructions: Two Guess and two Count instructions. The Guess instructions were intended to produce gaze behavior similar to free viewing with \daniel{fewer} task constraints than in the Count instructions that require identification of and search for objects. Contrary to free viewing, \daniel{however, under Guess instructions,} eye behavior across participants was expected to be \daniel{guided more strongly} by the same aspects of the image \daniel{to solve the tasks (e.g., shadows, daylight, vegetation)}. In the two Count conditions, participants needed to examine the entire image to detect and count all target objects. Thus, both Count tasks were considered as a form of search task as they included a search for target objects in an image.

We compared tasks similar to free viewing (Guess) with tasks similar to search (Count) by quantifying how well fixation locations in the four tasks were predicted by a recent saliency model \citep[DeepGaze2;][]{kummerer2016}. Since saliency models were designed to predict fixation locations during free viewing, we expected a better match between the predictions of the saliency model and the two free viewing tasks than the two search tasks \citep[cf.,][]{Schutt:2019}. Numerically, target selection in the Guess conditions was in better agreement with predictions from the saliency model than in the Count conditions. Statistically the predictions for the Guess \daniel{conditions} outperformed predictions of \daniel{the Count Animals condition}. The \daniel{Count People} condition lay \daniel{nearby the Guess conditions} and did \daniel{not} differ significantly from \daniel{these}.  Since saliency models typically incorporate detectors for persons and faces, a large fraction of fixations on persons and faces can be predicted in the Count People condition \citep[cf.,][]{Mackay.JVis.2012}. In summary, the Guess conditions resembled free viewing more than the Count conditions and, consequently, the Guess conditions generate eye movements similar to the free viewing instruction. \rev{It is important to note that the DeepGaze2 model included the central fixation tendency, so that the better prediction of the Guess conditions could be partly explained by the stronger central fixation bias in these conditions.}

Low predictive power of saliency models for fixation locations in search tasks has also been reported for the search of artificial targets embedded in scenes \citep{Rothkegel:2019vw,Schutt:2019} \daniel{as well as for searching images of real-world scenes for real-world objects \citep{henderson2007, foulsham2008b}}. While eye-movement parameters like fixation durations and saccade amplitudes adapted to the visibility of the target in the periphery \citep{Rothkegel:2019vw}, fixations were differently \daniel{associated} \rev{with} features in search and free viewing tasks. Even training a saliency model based on early visual processing to the data set did not improve predictions considerably \citep{Schutt:2019}. Our results demonstrate that the low predictive power of saliency models in the search tasks is also true for search tasks with \daniel{non-manipulated real-world scenes}. However, While fixation locations were not well predicted by the saliency model in the search tasks and in particular not in the Count Animals tasks, several other eye-movement parameters adapted to the search task. Fixation durations were shortest in the Count Animals condition, saccade amplitudes were shorter and the central fixation bias smaller in the Count conditions than the Guess conditions. Thus, there is no simple relation between low-level image features and fixation locations in search, but other parameters demonstrate that eye movements adapt to the specificities of the task. 

\subsection{Conclusions}
Due to several limitations the generalizability of theoretical implications of the scene-viewing paradigm has been criticized. However, real-world scenarios often lack experimental control and are detached from the previous research. Here we demonstrate that the advancements in mobile eye-tracking and image processing make it possible to deal with the limitations of the scene-viewing paradigm, while keeping high experimental control in a laboratory setup. \daniel{Our setup provides a fruitful, highly controlled but less constrained environment to investigate eye-movement control across tasks.}

\section{Acknowledgments}
\rev{We thank Benjamin W.~Tatler (Aberdeen) for valuable comments. This work was funded by the Deutsche Forschungsgemeinschaft through grants to Hans A. Trukenbrod (grant DFG TR 1385/2-1) \& Ralf Engbert (grant DFG EN 471/16-1).}

\bibliography{ms.bib}
\bibliographystyle{jovcite}

\section{Appendix 1}

\daniel{For our analyses, we used linear mixed effect models (LMM). For each dependent variable, we used the same fixed effect structure, except when explicitly stated otherwise (see Methods for details). For the random effect structure, we estimate random effects for participants and images. We first formulated the maximal possible random effect structure \citep{Barr:2013eh}, with random intercepts for images and participants and random slopes for each of our three contrasts. We reduced these models until the $lme4$-package returned no convergence problems. First, we removed correlation terms. Second, we removed the least varying random effect terms. If the reduced model converges, we try to re-include the correlation terms. In the case that we end up with two converging models of the same complexity, but with different terms for slopes in the image and participant random effect part, we used Bayesian-Information-Criterion (BIC) to determine the best model. We performed a principal component analysis to check whether all random effect terms explain non-zero variance; thus, none of the models was degenerate \citep{Bates:2015vla}.
The random effect structure for entropy differed from all others since the entropy measurement is based on fixations from all participants over images, we did not estimate random effects for participants.}

\begin{table}[H]
\begin{center}
\color{black}

\caption{Random effects structure.}
\begin{tabular}{lrcrlcrl}
\noalign{\vskip0.3cm}
\multicolumn{3}{l}{Dependent variable} &  \multicolumn{3}{c}{Random effect participant part} & \multicolumn{2}{c}{Random effect image part}  \\

\hline
\multicolumn{8}{l}{\textit{Fixation duration}}\\
DV$\sim$ & fixed effects & + & (1 + C1 & || participant) & + & (1 + C1 + C2 & || image) \\

\multicolumn{8}{l}{} \\
\multicolumn{8}{l}{\textit{Saccade amplitudes}}\\
DV$\sim$ & fixed effects & + & \multicolumn{1}{r}{(1 + C1 } & || participant) & + & \multicolumn{1}{r}{(1 + C1 + C2} & || image) \\

\multicolumn{8}{l}{} \\
\multicolumn{8}{l}{\textit{Central fixation bias*}}\\
DV$\sim$ & fixed effects & + & (1 & | participant) & + & (1 & | image) \\

\multicolumn{8}{l}{} \\
\multicolumn{8}{l}{\textit{Entropy}}\\
DV$\sim$ & fixed effects &   &   &   & + & (1 + C1 + C2 & || image) \\

\multicolumn{8}{l}{} \\
\multicolumn{8}{l}{\textit{Predictability}} \\
DV$\sim$ & fixed effects & + & (1 + C1 + C2 & || participant) & + & (1 + C1 + C2 + C3 & || image) \\

\multicolumn{8}{l}{} \\
\multicolumn{8}{l}{\textit{Predictability per task*}} \\
DV$\sim$ & fixed effects & + & (1  & | participant) & + & (1  & | image) \\

\multicolumn{8}{l}{} \\
\multicolumn{8}{l}{\textit{DeepGaze2}} \\
DV$\sim$ & fixed effects & + & (0 + C1 + C2 + C3 & || participant) & + & (1 + C1 + C2 + C3 & | image)
\\
\hline
\end{tabular}\\
Notes: 1 Intercept, C1 first contrast the two Guess against the two Count tasks, C2 second contrast Count Animals against Count People, C3 third contrast Guess Time against Guess Country, || zero correlation parameter, DV dependent variable, * we choose the minimal model with only random intercepts for participants and images to have comparable models between all subsets of this analysis.  
\label{ex:ranef}
\end{center}
\end{table}

\end{document}